\documentclass[twocolumn,reprint,secnumarabic,amssymb,amsmath,nobibnotes,aps,showpacs]{revtex4-1}

\usepackage{graphics} \usepackage{graphicx} \usepackage{braket} \usepackage{float} \usepackage{tabularx} 
\usepackage[hyperindex,breaklinks]{hyperref}
\usepackage[usenames,dvipsnames]{color}

\usepackage{braket} \usepackage{physics} 
\usepackage[hyperindex,breaklinks]{hyperref}

\usepackage{mathrsfs}
\usepackage{tabularx}

\usepackage{multirow}

\usepackage{rotating}

\begin{document}

\title{Nuclear-Spin Dependent Parity Violation in Optically Trapped Polyatomic Molecules }

\author{E. B. Norrgard, D. S. Barker, S. Eckel, J. A. Fedchak, N. N. Klimov, J. Scherschligt}
\affiliation{Sensor Sciences Division, National Institute of Standards and Technology, Gaithersburg, MD 20899, USA}
\email[Electronic address: ]{eric.norrgard@nist.gov}

\begin{abstract} We investigate using optically trapped linear polyatomic molecules as probes of nuclear spin-dependent parity violation.   The presence of closely spaced, opposite-parity $\ell$-doublets is a general feature of such molecules, allowing parity-violation-sensitive pairs of levels to be brought to degeneracy in magnetic fields typically 100 times smaller than in diatomics.    Assuming laser cooling and trapping of polyatomics at the current state-of-the-art for diatomics, we expect to measure nuclear spin-dependent parity-violating matrix elements $iW$ with 70 times better sensitivity than  the current best measurements.  Our scheme should allow for 10\,\% measurements of $iW$ in nuclei as light as Be or as heavy as Yb, with averaging times on order the of 10 days and 1 second, respectively.

\end{abstract}


\maketitle

Measurements of nuclear spin-independent (NSI) and nuclear spin-dependent (NSD)  parity violation (PV) are a means to probe Standard Model (SM) electroweak interactions on a tabletop scale \cite{Khriplovich1997}. NSI-PV has been measured in protons and a number of heavy atoms \cite{Qweak2018, Macpherson1991,Meekhof1993,Vetter1995,Nguyen1997,Tsigutkin2009} and found to be in good agreement with SM PV predictions  due to the weak charge $Q_W$.  However, the only non-zero measurement (14\,\% relative uncertainty) of NSD-PV in an atomic system comes from Cs \cite{Wood1997}, and this result implies constraints on SM meson-nucleon couplings which are in disagreement with other atomic PV measurements \cite{Haxton2001,Johnson2003}.
 NSD-PV arises primarily from two interactions: vector electron-axial nucleon electroweak current coupling ($V_eA_n$), and the nuclear anapole moment.  The $V_eA_n$ effect is described by two parameters $C_{2u}$ and $C_{2d}$ relating to spin-dependent $Z_0$ boson exchange between an electron and an up or down quark, respectively.  These parameters are among the most poorly measured  in the SM, with relative uncertainties 300\,\% and 70\,\%, respectively \cite{Wang2014}.  PV measurements may also probe beyond standard model physics \cite{Langacker1992,Dzuba2017}.  Searches for oscillating PV signals have been proposed as a means to detect axion-like particles, a leading dark matter candidate \cite{Stadnik2014}.



A beam of cold diatomic molecules has been demonstrated \cite{Kozlov1986,Flambaum1985,DeMille2008b,Altuntas2018, Altuntas2018b} to be a highly sensitive system for measuring NSD-PV effects.  Mixing of opposite-parity quantum states from PV effects is amplified when the states have nearly the same energy \cite{Nguyen1997}. The lowest two rotational states of diatomic molecules have opposite parity and may be brought to near degeneracy using a large magnetic field $\mathcal{B}$. While this method is quite general, current systematic uncertainties are roughly 100 times too large to measure NSD-PV in the lightest nuclei where nuclear structure calculations are tractable \cite{Maris2011}.

 Recent advances in laser cooling have lead to optically trapped diatomic molecules with sub-Doppler temperature and single-molecule detection efficiency \cite{Anderegg2018}, while similar strides  with polyatomic molecules have followed closely behind \cite{Prehn2016,Kozyryev2016,Kozyryev2016b,Kozyryev2017}.  Moreover, polyatomic molecules have been proposed as exquisite systems for precision measurements of fundamental symmetries \cite{Isaev2017,Kozyryev2018,Kozyryev2017b} and time variation of fundamental constants \cite{Kozlov2013,Prehn2018}.

Here, we show that linear asymmetric polyatomic molecules in an optical trap are well-suited for measurement of NSD-PV.
Polyatomic molecules possess opposite-parity states $10$ to 1000 times closer in energy than diatomics, requiring similarly smaller $\mathcal{B}$-fields.
  Systematic uncertainties are reduced compared to beam experiments due to the lower magnetic field and a smaller interaction volume.
   Furthermore, these smaller fields may be produced without superconducting magnets, allowing trivial $\mathcal{B}$-field reversal for detection and mitigation of systematic effects.  We show that these molecule may be optically trapped using ``magic'' conditions where differential light shifts are small enough for a precise PV measurement. The obvious advantage of performing a precision measurement on trapped species compared to a beam is the increased interaction time $\tau$.  The sensitivity to any PV matrix element $iW$ is $\delta W/W$\,=\,$1/\tau\sqrt{N_m}$, where $N_m$ is the total number of measurements ($iW$ is purely imaginary due to conservation of time-reversal symmetry).  Assuming optical trapping of polyatomic molecules at the current state-of-the-art for diatomics \cite{Park2017,Anderegg2018,Cheuk2018}, we expect at least a factor of 70 increase in PV sensitivity over the state-of-the-art  BaF measurement \cite{Cahn2014, Altuntas2018}.   Our method is applicable to all laser-coolable polyatomic molecules with $^2\Sigma$ ground states, and possibly others.

Consider the properties of linear asymmetric  molecules with a $^2\Sigma^+$ electronic ground state.
 If a bending vibrational mode (with vibrational constant $\omega_b$ and quantum number $v_b$) is excited and all other vibrational modes are in their ground state, the molecule's rotational angular momentum $\boldsymbol{N}$ has a projection along the molecular axis $\ell = \pm v_b, \pm (v_b-2), \ldots, \pm1$ or 0.
Within this vibrational manifold, the effective Hamiltonian is
\begin{equation}\label{eq:hamiltonian srhf}
\begin{aligned}
  H =& B (\boldsymbol{N}^2-\ell^2)\pm(-1)^N\frac{q_b}{2}\boldsymbol{N}^2 + \gamma \boldsymbol{N}\cdot\boldsymbol{S}+ b\boldsymbol{I}\cdot\boldsymbol{S}\\
  &\quad +c I_z S_z -e\rm{T}^2(\nabla\boldsymbol{E})\cdot\rm{T}^2(\boldsymbol{Q}) \\
  &\quad+(\mu_B g_S \boldsymbol{\mathcal{B}}\cdot\boldsymbol{S}+\mu_B g_L \boldsymbol{\mathcal{B}}\cdot\boldsymbol{L}+\mu_N g_I \boldsymbol{\mathcal{B}}\cdot\boldsymbol{I}),
\end{aligned}
\end{equation}
where $B$ is the rotational constant, $\gamma$ is the spin-rotation (SR) constant, $b$ and $c$ are hyperfine (HF) constants, $e$ is the electron charge, $\rm{T}^2(\nabla\boldsymbol{E})\cdot\rm{T}^2(\boldsymbol{Q})$ is a scalar product of rank-2 spherical tensors describing the electric field gradient $\nabla\boldsymbol{E}$ at the nucleus with quadrupole moment $\boldsymbol{Q}$, $\boldsymbol{S}$ is the electron spin, $\boldsymbol{L}$ is the electron orbital angular momentum, and $\boldsymbol{I}$ is the nuclear spin \cite{Hirota1985, Brown2003}. The upper (lower) sign corresponds to the parity $\mathcal{P}$ of the closely spaced ``$\ell$-doublet''  eigenstates \mbox{$\ket{N,{v_b}^{\ell}, \mathcal{P}=\pm} = \frac{1}{\sqrt{2}}(\ket{N,+\ell}\pm(-1)^{N-\ell}\ket{N,-\ell})$}.
The $\ell$-doublet is the key property of polyatomic molecules absent from diatomics which we wish to exploit for a PV measurement.  In linear modes, as well as diatomics, opposite-parity levels are spaced by roughly the rotational constant $B/2\pi$\,$\sim$\,$1$\,GHz to 100 GHz. In an excited bending mode, opposite-parity states are spaced by only $q_b$\,$\approx$\,$-2B^2/\omega_b$, which is on the order of $q_b/2\pi\sim 10$\,MHz to 100\,MHz.    The relative spacing between levels may be tuned to degeneracy via the Zeeman interaction (last three terms of Eq.\,\ref{eq:hamiltonian srhf}, with $\mu_B$, $\mu_N$ the Bohr and nuclear magneton, respectively; and $g_S$, $g_L$, and $g_I$ the $g$-factors corresponding to $\boldsymbol{S}$, $\boldsymbol{L}$, and $\boldsymbol{I}$, respectively).

For a given electronic state of a molecule, the effective NSD-PV Hamiltonian is \cite{Flambaum1980}
\begin{equation}\label{eq:effective hamiltonian}
 H_p^{\rm{eff}} = \kappa W_p (\boldsymbol{S}\times \boldsymbol{\hat{I}})\cdot \boldsymbol{\hat{n}}.
\end{equation}
Here, $W_p$ encodes the overlap of unpaired electrons with the nucleus and can be calculated from molecular spectroscopy with high precision for $^2\Sigma^+$ states \cite{Kozlov1986}, $\boldsymbol{\hat{n}}$ is a unit vector along the molecular axis, and
 $\kappa$ is the measurable parameter of interest. In a given nucleus, various NSD-PV effects contribute to $\kappa = \kappa_2 + \kappa_a + \kappa_Q$.  $\kappa_2$ is proportional to the strength of the $V_eA_n$ coupling, and is independent of nuclear mass $A$ (for a typical nucleus, $\vert\kappa_2\vert \simeq 0.05$).  $\kappa_a$ is proportional to the nuclear anapole moment and is proportional to $A^{2/3}$.  $\kappa_Q$ is due to the combined effects of nuclear weak charge and normal hyperfine structure, and is negligible compared to $\kappa_2$ and $\kappa_a$ \cite{Sheng2010}.  
 Measurements in several nuclei are required to distinguish among the different NSD-PV effects.

 Ultimately, the ability to precisely determine $\kappa$ and its underlying contributions  will be limited to the accuracy of theoretical values of $W_p$.    Calculations of $W_p$ have been performed on several diatomic molecules via Dirac-Hartree-Fock and relativistic density-functional \cite{Borschevsky2013}, quasirelativistic zero-order regular approximation \cite{Isaev2017}, and (with an estimated 1.5\,\% accuracy) relativistic coupled-cluster \cite{Hao2018} methods.  While such calculations are beyond the scope of this proposal, a semiempirical method may be used to calculate $W_p$ for any species to approximately 10\,\%, assuming the SR/HF constants in Eq.\,\ref{eq:hamiltonian srhf} are known \cite{Kozlov1986}.  The SR/HF constants relating to a typical metal atom M  differ by a approximately 10\,\% between laser-coolable monofluorides MF and MX molecules, where X is a suitable ligand with charge state -1 (e.g., OH, NC, or CCH).  Thus,  we expect similar 10\,\% accuracy when estimating $W_p$ for MX, using either MX SR/HF constants for a semiemprical calculation or a more detailed method described above for the corresponding MF.  In Table I, we give a list of laser-coolable polyatomic molecules MX, with $W_p$ from calculations on MX where available or MF otherwise.

\begin{table}[t]\label{tab:sensitivity1}
\centering
  \begin{tabularx}{.9\columnwidth}{r c c c r  c}
    \hline
    Species &$I$ &  $100\kappa_a$ & $100\kappa_2$ & $W_p/2\pi$ (Hz) &  $W^{m}/2\pi$ (Hz) \\
    \hline
    $^{9}$BeNC   &3/2 &$-0.66$  &-5.0  &0.46 \cite{Hao2018} & 0.010\\
    $^{25}$MgNC  &5/2 &-1.4  &-5.0  &4.91 \cite{Hao2018} & 0.13\\
    $^{43}$CaOH  &7/2 &-2.1  &-5.0  &10.8 \cite{Hao2018} & 0.31\\
    $^{87}$ SrOH &9/2 &-3.4  &-5.0  &51 \cite{Hao2018}   & 1.7\\
    $^{137}$BaOH &3/2 &4.1   &3.0   &147 \cite{Hao2018}  & 4.2\\
    $^{171}$YbOH &1/2 &3.9   &1.7   &576 \cite{Borschevsky2013}  &12.9\\
    $^{225}$RaOH &1/2 &-4.7  &-5.0  &1400 \cite{Isaev2017}  &54\\ \hline

  \end{tabularx}\caption{Parameters for PV measurement in polyatomic molecules.  Values of $W_p$ for MX are taken to be equal to that of the corresponding MF molecule given in Refs.\,\cite{Borschevsky2013, Hao2018}, except RaOH for which calculation exists \cite{Isaev2017}.  Values for $\kappa_2$ assume $C_{2N}$\,=\,$-0.05$; values for $\kappa_a$ assume weak neutron-neutron coupling $g$\,=\,1 \cite{khriplovich1991}.  $W^{m}$ is the value of $W$ with maximal $\ev{(\boldsymbol{S}\times\boldsymbol{\hat{I}})\cdot \boldsymbol{\hat{n}}}$. }
\end{table}

In general, Eq.\,\ref{eq:effective hamiltonian} should be summed over all nuclei $i$ with spin $\boldsymbol{I}_i \geq 1/2 $ in a molecule.  The PV signal is easiest to interpret when the unpaired electron is centered on one atom in the molecule, i.e.\ $W_p^{(i)}\approx 0$ for all but one atom. A single-atom-centered unpaired electron is also a defining characteristic of laser-coolable molecules:  this electron interacts negligibly with the nuclear vibration of the molecule, leading  to electronic transitions which are highly diagonal in vibrational quantum number (and thus requires a small number of vibrational repump lasers for cooling) \cite{DiRosa2004}.  Laser cooling schemes have been proposed for molecules with an electron centered on atoms with a wide range of mass (as light as Be \cite{Lane2012} and B \cite{Hendricks2014}, and as heavy as Yb\cite{Lim2018} and Tl \cite{Hunter2012,Norrgard2017}), and extension to polyatomic species, while technically more complicated, is straightforward \cite{Kozyryev2016b}.

$H_p^{\rm{eff}}$ is a pseudoscalar interaction, which connects states with different parity $\mathcal{P}$ and the same lab frame angular momentum projection $m_F$.  We dub such states $\ket{\tilde{\eta}; m_F,+},\ket{\tilde{\eta}^\prime; m_F,-}$ a ``PV pair'', with $\tilde{\eta}$ denoting all other nominal quantum numbers when $\mathcal{B}$\,=\,0.
So long as the $\ell$-doublet splitting  is not smaller than all SR/HF splittings, PV pairs of a given rotational manifold $N$ cross in an applied $\mathcal{B}$-field when $\mu_B \mathcal{B}$\,$\sim$\,$q_b$.  This situation is common for light molecules due to their typically smaller HF interactions and larger $q_b$ arising from their larger rotational constant $B$.  Typical values of $q_b$ imply a modest field of $\mathcal{B}$\,$\sim$\,1\,mT to 10\,mT will bring a PV pair to degeneracy.
For instances where  the $\ell$-doublet splitting is smaller than the SR/HF splitting, states $\ket{\tilde{\eta};m_F,\pm}$ remain split by $\sim$\,$q_b$ for any applied $\mathcal{B}$-field. In this case, the Zeeman interaction repels this $\ell$-doublet from a neighboring $\ket{\tilde{\eta}^\prime;m_F,\pm}$ doublet, preventing the crossing of PV pairs $\ket{\tilde{\eta};m_F,\pm},\ket{\tilde{\eta}^\prime;m_F,\mp}$.   The first PV crossing actually occurs at $\mu_B \mathcal{B}$\,$\approx$\,$B$, as in a diatomic molecule \cite{DeMille2008b}, and requires an experimentally more challenging $\mathcal{B}$\,$\sim$\,100\,mT to 1000\,mT \cite{DeMille2008b}.



In some cases, it may be advantageous to measure $iW$ in an excited rotational or vibrational state.  For example, consider $^{171}$YbOH ($\boldsymbol{I}$($^{171}$Yb)\,=\,1/2). We may estimate the relevant parameters in Eq.\,\ref{eq:hamiltonian srhf} by reduced-mass-scaling (where appropriate) the constants $B$, $\omega_{b=2}$, $\gamma$, $b$(H), and $c$(H) from the $^{174}$YbOH isotopologue ($\boldsymbol{I}$($^{174}$Yb)\,=\,0) \cite{Nakhate2018}; by assuming constants $b$(Yb) and $c$(Yb) to be the same as in chemically similar $^{171}$YbF; and by
taking $q_b=-2B^2/\omega_2$.
With these parameters, the $\ell$-doublet in $\ket{{v_b}^\ell=1^1, N=1}$ is smaller than the SR and Yb HF interactions.  For all possible PV pairs, the value of $\ev{(\boldsymbol{S}\times \boldsymbol{\hat{I}}(\rm{Yb}))\cdot \boldsymbol{\hat{n}}}$ is only nonzero due to the small state mixing from the HF interaction with the H nucleus. However, in $\ket{{v_b}^\ell=1^1, N=2,3}$, the $\ell$-doublet is larger than the Yb HF splitting and multiple PV pairs with $\ev{(\boldsymbol{S}\times \boldsymbol{\hat{I}}(\rm{Yb}))\cdot \boldsymbol{\hat{n}}}$\,$\sim$\,0.1 exist.

We now propose a procedure to trap polyatomic molecules and measure matrix elements $iW$ of $H_p^{\rm{eff}}$.
 A cyrogenic buffer gas beam source creates a slow, cold beam of the desired molecule species  \cite{Maxwell2005, Hutzler2012}.  Molecules are sequentially laser-slowed \cite{Barry2012},  trapped using a magneto-optical trap (MOT) \cite{Barry2014},  and loaded into a red-detuned optical dipole trap (ODT) while performing $\Lambda$-enhanced cooling \cite{Cheuk2018}.  Then, one of the cooling laser frequencies is turned off in order to optically pump molecules into a single, optically dark SR/HF state.
 Stimulated Raman adiabatic passage completes state preparation by efficiently transferring to the $\ket{{v_b}^\ell=1^1,N,\mathcal{P}=(-1)^N}$ state \cite{Chotia2012, Panda2016}.

The PV signal is measured by the Stark interference method, which has been examined in detail elsewhere \cite{Nguyen1997,DeMille08b,Tsigutkin2009}. We summarize the main points closely following the notation of Ref.\,\cite{Cahn2014}.  We apply a static magnetic field $\boldsymbol{\mathcal{B}}$\,=\,$\mathcal{B}\hat{\boldsymbol{z}}$ to  shift a particular PV pair to near degeneracy. We denote the time-dependent probability amplitudes of these states $c_\pm(t)$, and assume an initial state $c_-(0)$\,=\,1, $c_+(0)$\,=\,0.  An oscillating electric field $\boldsymbol{\mathcal{E}} = \mathcal{E}_0 \cos(\omega_\mathcal{E} t) \hat{\boldsymbol{z}}$ is applied to drive the transition between the near degenerate levels.
 The effective Hamiltonian $H_\pm^{\rm{eff}}$ for the two level system can be written \cite{Cahn2014}
\begin{equation}\label{eq:tls}
  H_\pm^{\rm{eff}} = \left( \begin{matrix}
                              \Delta & d\,\mathcal{E}_0\,\text{cos}(\omega_\mathcal{E}\,t) + i\,W \\
                              d\,\mathcal{E}_0\,\text{cos}(\omega_\mathcal{E}\,t) - i\,W & -\alpha^\prime\,{\mathcal{E}_0}^2\cos^2(\omega_\mathcal{E}\,t)/2
                            \end{matrix} \right).
\end{equation}
Here $\Delta$ is the small detuning from degeneracy, $d$ is the transition dipole moment, and $\alpha^\prime$ is the differential polarizability of the two states.   In the limit where  $W \ll d\,\mathcal{E}_0, \Delta \ll \omega_\mathcal{E}$, and assuming for now that $\alpha^\prime = 0$, the PV signal $S$\,=\,$\vert c_+(t) \vert^2 $ is
\begin{equation}\label{eq:probability amplitude}
  S \approx 4 \left[2\frac{W}{\Delta}\frac{d\,\mathcal{E}_0}{\omega_\mathcal{E}}+\left(\frac{d\,\mathcal{E}_0}{\omega_\mathcal{E}}\right)^2\right] \text{sin}^2 \left(\frac{\Delta\,t}{2}\right).
\end{equation}
From Eq.\,\ref{eq:probability amplitude}, we see that the PV transition amplitude (first term in square brackets) interferes  with the E1 transition amplitude (second term in square brackets).  The interference term changes sign under a reversal of either $\boldsymbol{\mathcal{E}}$, $\mathcal{B}$, or $\Delta$.  The PV matrix element $W$ may be extracted through an asymmetry measurement \cite{DeMille2008b}
\begin{equation}\label{eq:asymmetry}
  \mathcal{A}= \frac{S(+\mathcal{E}_0)-S(-\mathcal{E}_0)}{S(+\mathcal{E}_0)+S(-\mathcal{E}_0)} = 2 \frac{W}{\Delta} \frac{\omega_\mathcal{E}}{d\,\mathcal{E}_0}+ \ldots,
\end{equation}
where the ellipse denotes higher order terms in $W/\Delta$.
Detection using optical cycling should provide shot noise-limited readout.

\begin{figure}
   \centering
   \includegraphics[width= \columnwidth]{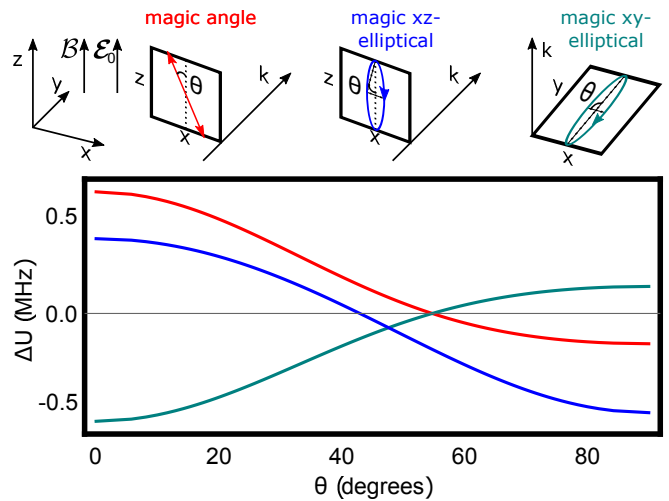}
   \caption{(Color online) Differential ac Stark shift $\Delta U$ as a function of polarization angle $\theta$ as defined in the figures for the nominal $\ket{N=1,J=1/2,F=3,m_F=3,\mathcal{P}=+}$, $\ket{N=1,J=3/2,F=3,m_F=3,\mathcal{P}=-}$ PV pair in $^{25}$MgNC. For each configuration $U$\,=\,1.2\,MHz for the PV pair when $\Delta U$\,=\,0 at $\theta_{\rm{magic}}$. The calculation is performed with $\mathcal{B}$\,$\approx$\,8.072\,mT such that the PV pair is degenerate in the absence of the trapping field.  (Red) linear polarization in the $x,z$-plane, (blue)  elliptical polarization in $x,z$-plane, (green) elliptical polarization in $x,y$-plane.  } \label{fig:magic}
 \end{figure}

 Sensitivity to $W$ is expected to be limited by homogeneous broadening of $\Delta$.  For optically trapped molecules, the detuning uncertainty $\delta\Delta$ is  dominated by the differential stark shift due to the optical trap.  Employing certain ``magic'' polarization conditions in an ODT \cite{Romalis1999,Kotochigova2010,Neyenhuis2012,Kim2013,Rosenband2018} will set $\alpha^\prime$\,=\,0, and $\delta\Delta$\,=\,$\delta U$\,=\,$\frac{\partial U}{\partial \theta}\delta\theta$, where $U$ is the trap depth and $\theta$ is an angle related to the polarization.   We investigated 3 such magic conditions for PV pairs in MgNC  and YbOH and find  $\frac{\partial U}{\partial \theta}\simeq U \delta \theta$ is typical.   Fig.\,\ref{fig:magic} demonstrates magic conditions for one PV pair in MgNC.  In producing magic conditions with a general elliptical polarization, we estimate $\delta \theta/\theta \sim 10^{-3}$ \cite{Kim2013}, while for linear polarized light, $\delta \theta/\theta < 10^{-4}$ is possible using high-quality Glan-type polarizers.  We therefore expect $\delta U/U$\,$\sim$\,10$^{-4}$ using a magic angle trap \cite{Romalis1999,Kotochigova2010}, where the ODT linear polarization is rotated by $\theta_{\rm{magic}}=\rm{cos}^{-1}(\frac{1}{\sqrt{3}})\approx54.7\,^\circ$ with respect to the quantization axis. For $U$\,=\,$2\pi$\,$\times$\,1\,MHz, $\delta\Delta$=$2\pi$\,$\times$\,100\,Hz.


An inhomogeneous $\mathcal{B}$-field will also produce broadening; typically $\delta\Delta$\,$\approx$\,$\mu_B\,\delta\mathcal{B}$ \cite{Altuntas2018b}.  We require  $\delta \mathcal{B} \ll 1/\delta U \mu_B$\,$\lesssim$\,1\,nT  for $\mathcal{B}$-field inhomogeneity to not limit sensitivity.  This implies $\delta\mathcal{B}/\mathcal{B}$\,=\,$10^{-7}$ for the largest fields we may require, $\mathcal{B}$\,=\,10\,mT.  For comparison, the recent BaF NSD-PV measurement demonstrated $\delta\mathcal{B}/\mathcal{B}$\,=\,$10^{-8}$, even with the much more experimentally challenging $\mathcal{B}$\,$\approx$\,460\,mT \cite{Altuntas2018,Altuntas2018b}.    We expect that the ability to easily reverse a smaller field will further aid in detecting and eliminating $\mathcal{B}$-field inhomogeneities. Moreover, the field must only be homogenous over the interaction volume, which is approximately 100 times smaller in an ODT compared a beam.  Systematic uncertainties involving field gradients should similarly by reduced a factor of 100 from the smaller interaction volume.

From Eq.\,\ref{eq:effective hamiltonian}, we see that Stark and NSD-PV amplitudes are $\pi/2$ out of phase, and there is no interference in a static $\mathcal{E}$-field.  However, the presence of a non-reversing $\mathcal{E}$-field $\mathcal{E}_{\rm{nr}}$ still poses an issue.  In the molecule frame, a static $\mathcal{E}_{\rm{nr}}$ has significant frequency components at axial and radial ODT frequencies $\omega_z, \omega_r$, and multiples, sums, and differences thereof.  Assuming uncorrelated trap oscillations, this effect will lead to a homogeneous broadening much smaller than that of the differential ac Stark shift.
     Accurate measurement of $\mathcal{E}_{\rm{nr}}$ is possible by Stark interference with a reversible pulsed dc field $\mathcal{E}_{\rm{r}}$  \cite{Altuntas2018b}, or by  microwave depletion spectroscopy \cite{OlearyThesis}.  Finally, investigating multiple PV pairs in the same molecule provides strong systematic error rejection by varying the ratio $\ev{(\boldsymbol{S}\times\boldsymbol{\hat{I}})\cdot \boldsymbol{\hat{n}}}/d$ by a calculable, possibly large, amount \cite{DeMille08b}. For example, in PV pairs with different signs of $\ev{(\boldsymbol{S}\times\boldsymbol{\hat{I}})\cdot \boldsymbol{\hat{n}}}/d$, contributions to $\mathcal{A}$ from actual NSD-PV will switch sign, but  contributions from $\mathcal{E}_{\rm{nr}}$ will not.

Other relaxation mechanisms are expected to lead to negligible broadening compared to differential ac Stark shifts.  For example, in a beam experiment, $\delta\Delta$ is typically limited by interaction time $\tau$, with $\tau$\,$\sim$\,100\,$\mu$s \cite{Altuntas2018}.  In an ODT, $\tau$ may easily exceed 1\,s.  Trapped molecule lifetimes $\tau_{\rm{trap}}$\,=\,0.5\,s to 25\,s have been reported in a variety of traps \cite{Norrgard2016, Anderegg2018,Williams2018,Chotia2012}.   With near ideal vacuum conditions, we expect trap lifetimes  $\tau_{\rm{trap}}$\,$\sim$\,10\,s, limited by vibrational decay.   The loss rate due to off-resonant scattering from the trapping laser  can typically be made $R_{\rm{sc}}$\,$\lesssim$\,1\,s$^{-1}$ by using standard mid-infrared wavelength fiber lasers.  Typical inelastic collision cross sections are expected to be $\sigma_{\rm{in}}$\,$\lesssim$\,$10^{-9}$\,cm$^3$/s.
Comparing with the trapping conditions of Ref.\,\cite{Cheuk2018} with $N$\,=\,1300 molecules at density $n$\,=\,$6$\,$\times$\,$ 10^8$\,cm$^{-3}$, we estimate an inelastic molecule-molecule collision rate of $R_{\rm{in}}$\,$\approx$\,0.5\,s$^{-1}$.  Therefore, collisions will become important when trapped molecule number $N$\,$\gtrsim$\,$10^5$, or with additional cooling. 

We now estimate the sensitivity of our method to NSD-PV matrix elements, $\delta W = 1/\tau\sqrt{\mathcal{R} N T}$, where $\mathcal{R}$ is the repetition rate, $N$ the number of trapped molecules per measurement, and $T$ the total measurement time.   We assume molecules are trapped in a $U$\,=\,$2\pi$\,$\times$\,1\,MHz deep magic angle trap, with $\delta \theta/\theta$\,=\,$10^{-4}$.  The combined effects of all relaxation times considered should allow for interaction times of $\tau$\,$\lesssim$\,1\,s, but peak sensitivity is achieved with $\tau$\,=\,$1/\delta\Delta$\,$\approx$\,$1/(U\delta\theta)$\,$\approx$\,1.6\,ms.
Allowing $t_{\rm{MOT}}$\,=\,50\,ms to load the MOT and $t_{\rm{trans}}$\,=\,40\,ms for state transfer, repetition rates $\mathcal{R}$\,=\,10\,s$^{-1}$ should be possible.  We expect that with molecules produced from an isotopically enriched source,  $N$\,$\approx$\,1000 for all species; this would be equivalent to the best to-date sample of directly cooled molecules in an ODT \cite{Cheuk2018}. Under these conditions, our expected experimental sensitivity is \mbox{$\delta W$\,$\approx$\,$2\pi$\,$\times$\,1\,Hz$/\sqrt{\rm{Hz}}$}.  This represents a factor of 70 improvement over the best to-date NSD-PV measurement in BaF \cite{Cahn2014,Altuntas2018,Altuntas2018b}.
 More ambitiously, one could plausibly expect $N$\,=\,$10^5$ to $10^6$ could be achieved with improved loading and cooling efficiency \cite{DeMarco2018,Kozyryev2017b}.

  With the proposed sensitivity, it should be possible to separate contributions to $\kappa$ from nuclear anapole ($\kappa_a \propto A^{2/3}$) and $V_eA_n$ ($\kappa_2$ $A$-independent) effects by measuring NSD-PV in a variety of nuclei. With the only non-zero NSD-PV measurement to-date in the heavy $^{133}$Cs \cite{Wood1997}, a precise measurement of PV in a light system would be especially illuminating.  In light nuclei, $\kappa$\,$\approx$\,$\kappa_2$.  As stated, $\kappa_2$ depends upon $C_{2u,d}$ which are among the most poorly know SM parameters and are suppressed at tree level.  Thus, a precise measurement of NSD-PV in light systems could potentially be sensitive to beyond SM physics above the $1$ TeV scale \cite{Langacker1992}.  In molecules such as BeNC and MgNC  the nuclear and molecular calculations are highly tractable.  Furthermore, $W_p$(N)\,$\sim$\,$W_p$(C)\,$\sim$\,$2\pi$\,$\times$\,$10$\,mHz in these systems;  a single species could provide a 10\,\% measurement of $\kappa$ for three nuclei ($^{13}$C, $^{14}$N, and either $^{9}$Be, $^{10}$Be, or $^{25}$Mg) with $T$\,$\lesssim$\,100 hours per nucleus.  The nuclear structure of $^{14}$N is of special interest and well studied due to the anomalously long  $^{14}$C$\rightarrow$$^{14}$N half-life and their role in radiocarbon dating  \cite{Maris2011}.

 Because $iW$ is enhanced by $\approx$\,$Z^2A^{2/3}$ in heavy species \cite{Borschevsky2013}, RaOH  appears especially promising \cite{Isaev2017}.  However, the high mass is a hinderance to effective laser-slowing by the standard methods for molecules \cite{Barry2012, Hutzler2012}. Moreover, the longest-lived Ra isotope possessing nuclear spin ($^{225}$Ra) has a half-life of only 15 days.  Nevertheless, MOTs of atomic $^{225}$Ra with typical $N$\,=\,1000 have been produced for atomic electric dipole measurements \cite{Parker2015};  with our expected sensitivity, only $N$\,=\,300 \textit{total} molecules would need to be detected for $\delta W/W$\,=\,0.1.


We have shown that optically trapped polyatomic molecules offer a dramatically enhanced sensitivity to parity-violating effects and additional checks of systematic errors.   Restriction to laser-coolable species still allows for measurement of NSD-PV in nuclei with a wide range of masses, necessary for determination of key SM parameters and tests of beyond SM physics.  The improved sensitivity should enable measurements of NSD-PV even in light nuclei where calculations are highly accurate.

Here, we have only considered linear asymmetric molecules.  Symmetric top molecules possess $k$-doublets of opposite parity (similar to $\ell$-doublets) even in their vibrational ground state.  For most $^2A_1$ states (analogue of $^2\Sigma$), the $k$-doublet splitting ($\sim$\,10\,kHz) is smaller than SR/HF and thus not suitable for Zeeman tuning PV pairs to degeneracy for a NSD-PV measurement.  However, the close spacing and miniscule differential Zeeman and ac Stark shifts of $k$-doublets may make symmetric top molecules in a magnetic or optical trap ideal for measuring NSI-PV.

\begin{acknowledgments}
The authors thank E. Altunta\ifmmode \mbox{\c{s}}\else \c{s}\fi{}, D. DeMille, N. Hutzler, and I. Kozyryev for insightful conversations into polyatomic molecular structure and PV measurements.  We thank E. Altunta\ifmmode \mbox{\c{s}}\else \c{s}\fi{} and E. Shirley for their careful reading of the manuscript.  EBN and DSB acknowledge support from the National Research Council Postdoctoral Research Associateship Program.
\end{acknowledgments}
\bibliography{thebib}

\begin{thebibliography}{64}%
\makeatletter
\providecommand \@ifxundefined [1]{%
 \@ifx{#1\undefined}
}%
\providecommand \@ifnum [1]{%
 \ifnum #1\expandafter \@firstoftwo
 \else \expandafter \@secondoftwo
 \fi
}%
\providecommand \@ifx [1]{%
 \ifx #1\expandafter \@firstoftwo
 \else \expandafter \@secondoftwo
 \fi
}%
\providecommand \natexlab [1]{#1}%
\providecommand \enquote  [1]{``#1''}%
\providecommand \bibnamefont  [1]{#1}%
\providecommand \bibfnamefont [1]{#1}%
\providecommand \citenamefont [1]{#1}%
\providecommand \href@noop [0]{\@secondoftwo}%
\providecommand \href [0]{\begingroup \@sanitize@url \@href}%
\providecommand \@href[1]{\@@startlink{#1}\@@href}%
\providecommand \@@href[1]{\endgroup#1\@@endlink}%
\providecommand \@sanitize@url [0]{\catcode `\\12\catcode `\$12\catcode
  `\&12\catcode `\#12\catcode `\^12\catcode `\_12\catcode `\%12\relax}%
\providecommand \@@startlink[1]{}%
\providecommand \@@endlink[0]{}%
\providecommand \url  [0]{\begingroup\@sanitize@url \@url }%
\providecommand \@url [1]{\endgroup\@href {#1}{\urlprefix }}%
\providecommand \urlprefix  [0]{URL }%
\providecommand \Eprint [0]{\href }%
\providecommand \doibase [0]{http://dx.doi.org/}%
\providecommand \selectlanguage [0]{\@gobble}%
\providecommand \bibinfo  [0]{\@secondoftwo}%
\providecommand \bibfield  [0]{\@secondoftwo}%
\providecommand \translation [1]{[#1]}%
\providecommand \BibitemOpen [0]{}%
\providecommand \bibitemStop [0]{}%
\providecommand \bibitemNoStop [0]{.\EOS\space}%
\providecommand \EOS [0]{\spacefactor3000\relax}%
\providecommand \BibitemShut  [1]{\csname bibitem#1\endcsname}%
\let\auto@bib@innerbib\@empty
\bibitem [{\citenamefont {Khriplovich}\ and\ \citenamefont
  {Lamoreaux}(1997)}]{Khriplovich1997}%
  \BibitemOpen
  \bibfield  {author} {\bibinfo {author} {\bibfnamefont {I.~B.}\ \bibnamefont
  {Khriplovich}}\ and\ \bibinfo {author} {\bibfnamefont {S.~K.}\ \bibnamefont
  {Lamoreaux}},\ }\href@noop {} {\emph {\bibinfo {title} {{C}{P} violation
  without strangeness: electric dipole moments of particles, atomics, and
  molecules}}}\ (\bibinfo  {publisher} {Springer-Verlag},\ \bibinfo {year}
  {1997})\BibitemShut {NoStop}%
\bibitem [{\citenamefont {Collaboration}(2018)}]{Qweak2018}%
  \BibitemOpen
  \bibfield  {author} {\bibinfo {author} {\bibfnamefont {T.~J. L.~Q.}\
  \bibnamefont {Collaboration}},\ }\href {\doibase 10.1038/s41586-018-0096-0}
  {\bibfield  {journal} {\bibinfo  {journal} {Nature}\ }\textbf {\bibinfo
  {volume} {557}},\ \bibinfo {pages} {207} (\bibinfo {year}
  {2018})}\BibitemShut {NoStop}%
\bibitem [{\citenamefont {Macpherson}\ \emph {et~al.}(1991)\citenamefont
  {Macpherson}, \citenamefont {Zetie}, \citenamefont {Warrington},
  \citenamefont {Stacey},\ and\ \citenamefont {Hoare}}]{Macpherson1991}%
  \BibitemOpen
  \bibfield  {author} {\bibinfo {author} {\bibfnamefont {M.~J.~D.}\
  \bibnamefont {Macpherson}}, \bibinfo {author} {\bibfnamefont {K.~P.}\
  \bibnamefont {Zetie}}, \bibinfo {author} {\bibfnamefont {R.~B.}\ \bibnamefont
  {Warrington}}, \bibinfo {author} {\bibfnamefont {D.~N.}\ \bibnamefont
  {Stacey}}, \ and\ \bibinfo {author} {\bibfnamefont {J.~P.}\ \bibnamefont
  {Hoare}},\ }\href {\doibase 10.1103/PhysRevLett.67.2784} {\bibfield
  {journal} {\bibinfo  {journal} {Phys. Rev. Lett.}\ }\textbf {\bibinfo
  {volume} {67}},\ \bibinfo {pages} {2784} (\bibinfo {year}
  {1991})}\BibitemShut {NoStop}%
\bibitem [{\citenamefont {Meekhof}\ \emph {et~al.}(1993)\citenamefont
  {Meekhof}, \citenamefont {Vetter}, \citenamefont {Majumder}, \citenamefont
  {Lamoreaux},\ and\ \citenamefont {Fortson}}]{Meekhof1993}%
  \BibitemOpen
  \bibfield  {author} {\bibinfo {author} {\bibfnamefont {D.~M.}\ \bibnamefont
  {Meekhof}}, \bibinfo {author} {\bibfnamefont {P.}~\bibnamefont {Vetter}},
  \bibinfo {author} {\bibfnamefont {P.~K.}\ \bibnamefont {Majumder}}, \bibinfo
  {author} {\bibfnamefont {S.~K.}\ \bibnamefont {Lamoreaux}}, \ and\ \bibinfo
  {author} {\bibfnamefont {E.~N.}\ \bibnamefont {Fortson}},\ }\href {\doibase
  10.1103/PhysRevLett.71.3442} {\bibfield  {journal} {\bibinfo  {journal}
  {Phys. Rev. Lett.}\ }\textbf {\bibinfo {volume} {71}},\ \bibinfo {pages}
  {3442} (\bibinfo {year} {1993})}\BibitemShut {NoStop}%
\bibitem [{\citenamefont {Vetter}\ \emph {et~al.}(1995)\citenamefont {Vetter},
  \citenamefont {Meekhof}, \citenamefont {Majumder}, \citenamefont
  {Lamoreaux},\ and\ \citenamefont {Fortson}}]{Vetter1995}%
  \BibitemOpen
  \bibfield  {author} {\bibinfo {author} {\bibfnamefont {P.~A.}\ \bibnamefont
  {Vetter}}, \bibinfo {author} {\bibfnamefont {D.~M.}\ \bibnamefont {Meekhof}},
  \bibinfo {author} {\bibfnamefont {P.~K.}\ \bibnamefont {Majumder}}, \bibinfo
  {author} {\bibfnamefont {S.~K.}\ \bibnamefont {Lamoreaux}}, \ and\ \bibinfo
  {author} {\bibfnamefont {E.~N.}\ \bibnamefont {Fortson}},\ }\href {\doibase
  10.1103/PhysRevLett.74.2658} {\bibfield  {journal} {\bibinfo  {journal}
  {Phys. Rev. Lett.}\ }\textbf {\bibinfo {volume} {74}},\ \bibinfo {pages}
  {2658} (\bibinfo {year} {1995})}\BibitemShut {NoStop}%
\bibitem [{\citenamefont {Nguyen}\ \emph {et~al.}(1997)\citenamefont {Nguyen},
  \citenamefont {Budker}, \citenamefont {DeMille},\ and\ \citenamefont
  {Zolotorev}}]{Nguyen1997}%
  \BibitemOpen
  \bibfield  {author} {\bibinfo {author} {\bibfnamefont {A.~T.}\ \bibnamefont
  {Nguyen}}, \bibinfo {author} {\bibfnamefont {D.}~\bibnamefont {Budker}},
  \bibinfo {author} {\bibfnamefont {D.}~\bibnamefont {DeMille}}, \ and\
  \bibinfo {author} {\bibfnamefont {M.}~\bibnamefont {Zolotorev}},\ }\href
  {\doibase 10.1103/PhysRevA.56.3453} {\bibfield  {journal} {\bibinfo
  {journal} {Phys. Rev. A}\ }\textbf {\bibinfo {volume} {56}},\ \bibinfo
  {pages} {3453} (\bibinfo {year} {1997})}\BibitemShut {NoStop}%
\bibitem [{\citenamefont {Tsigutkin}\ \emph {et~al.}(2009)\citenamefont
  {Tsigutkin}, \citenamefont {Dounas-Frazer}, \citenamefont {Family},
  \citenamefont {Stalnaker}, \citenamefont {Yashchuk},\ and\ \citenamefont
  {Budker}}]{Tsigutkin2009}%
  \BibitemOpen
  \bibfield  {author} {\bibinfo {author} {\bibfnamefont {K.}~\bibnamefont
  {Tsigutkin}}, \bibinfo {author} {\bibfnamefont {D.}~\bibnamefont
  {Dounas-Frazer}}, \bibinfo {author} {\bibfnamefont {A.}~\bibnamefont
  {Family}}, \bibinfo {author} {\bibfnamefont {J.~E.}\ \bibnamefont
  {Stalnaker}}, \bibinfo {author} {\bibfnamefont {V.~V.}\ \bibnamefont
  {Yashchuk}}, \ and\ \bibinfo {author} {\bibfnamefont {D.}~\bibnamefont
  {Budker}},\ }\href {\doibase 10.1103/PhysRevLett.103.071601} {\bibfield
  {journal} {\bibinfo  {journal} {Phys. Rev. Lett.}\ }\textbf {\bibinfo
  {volume} {103}},\ \bibinfo {pages} {071601} (\bibinfo {year}
  {2009})}\BibitemShut {NoStop}%
\bibitem [{\citenamefont {Wood}\ \emph {et~al.}(1997)\citenamefont {Wood},
  \citenamefont {Bennett}, \citenamefont {Cho}, \citenamefont {Masterson},
  \citenamefont {Roberts}, \citenamefont {Tanner},\ and\ \citenamefont
  {Wieman}}]{Wood1997}%
  \BibitemOpen
  \bibfield  {author} {\bibinfo {author} {\bibfnamefont {C.~S.}\ \bibnamefont
  {Wood}}, \bibinfo {author} {\bibfnamefont {S.~C.}\ \bibnamefont {Bennett}},
  \bibinfo {author} {\bibfnamefont {D.}~\bibnamefont {Cho}}, \bibinfo {author}
  {\bibfnamefont {B.~P.}\ \bibnamefont {Masterson}}, \bibinfo {author}
  {\bibfnamefont {J.~L.}\ \bibnamefont {Roberts}}, \bibinfo {author}
  {\bibfnamefont {C.~E.}\ \bibnamefont {Tanner}}, \ and\ \bibinfo {author}
  {\bibfnamefont {C.~E.}\ \bibnamefont {Wieman}},\ }\href@noop {} {\bibfield
  {journal} {\bibinfo  {journal} {Science}\ }\textbf {\bibinfo {volume}
  {275}},\ \bibinfo {pages} {1759} (\bibinfo {year} {1997})}\BibitemShut
  {NoStop}%
\bibitem [{\citenamefont {Haxton}\ and\ \citenamefont
  {Wieman}(2001)}]{Haxton2001}%
  \BibitemOpen
  \bibfield  {author} {\bibinfo {author} {\bibfnamefont {W.~C.}\ \bibnamefont
  {Haxton}}\ and\ \bibinfo {author} {\bibfnamefont {C.~E.}\ \bibnamefont
  {Wieman}},\ }\href@noop {} {\bibfield  {journal} {\bibinfo  {journal} {Annual
  Review of Nuclear and Particle Science}\ }\textbf {\bibinfo {volume} {51}},\
  \bibinfo {pages} {261} (\bibinfo {year} {2001})}\BibitemShut {NoStop}%
\bibitem [{\citenamefont {Johnson}\ \emph {et~al.}(2003)\citenamefont
  {Johnson}, \citenamefont {Safronova},\ and\ \citenamefont
  {Safronova}}]{Johnson2003}%
  \BibitemOpen
  \bibfield  {author} {\bibinfo {author} {\bibfnamefont {W.~R.}\ \bibnamefont
  {Johnson}}, \bibinfo {author} {\bibfnamefont {M.~S.}\ \bibnamefont
  {Safronova}}, \ and\ \bibinfo {author} {\bibfnamefont {U.~I.}\ \bibnamefont
  {Safronova}},\ }\href {\doibase 10.1103/PhysRevA.67.062106} {\bibfield
  {journal} {\bibinfo  {journal} {Phys. Rev. A}\ }\textbf {\bibinfo {volume}
  {67}},\ \bibinfo {pages} {062106} (\bibinfo {year} {2003})}\BibitemShut
  {NoStop}%
\bibitem [{\citenamefont {{The Jefferson Lab PVDIS Collaboration, Wang
  \textit{et al.}}}(2014)}]{Wang2014}%
  \BibitemOpen
  \bibfield  {author} {\bibinfo {author} {\bibnamefont {{The Jefferson Lab
  PVDIS Collaboration, Wang \textit{et al.}}}},\ }\href
  {http://dx.doi.org/10.1038/nature12964} {\bibfield  {journal} {\bibinfo
  {journal} {Nature}\ }\textbf {\bibinfo {volume} {506}},\ \bibinfo {pages} {67
  EP } (\bibinfo {year} {2014})}\BibitemShut {NoStop}%
\bibitem [{\citenamefont {Langacker}\ \emph {et~al.}(1992)\citenamefont
  {Langacker}, \citenamefont {Luo},\ and\ \citenamefont
  {Mann}}]{Langacker1992}%
  \BibitemOpen
  \bibfield  {author} {\bibinfo {author} {\bibfnamefont {P.}~\bibnamefont
  {Langacker}}, \bibinfo {author} {\bibfnamefont {M.}~\bibnamefont {Luo}}, \
  and\ \bibinfo {author} {\bibfnamefont {A.~K.}\ \bibnamefont {Mann}},\ }\href
  {\doibase 10.1103/RevModPhys.64.87} {\bibfield  {journal} {\bibinfo
  {journal} {Rev. Mod. Phys.}\ }\textbf {\bibinfo {volume} {64}},\ \bibinfo
  {pages} {87} (\bibinfo {year} {1992})}\BibitemShut {NoStop}%
\bibitem [{\citenamefont {Dzuba}\ \emph {et~al.}(2017)\citenamefont {Dzuba},
  \citenamefont {Flambaum},\ and\ \citenamefont {Stadnik}}]{Dzuba2017}%
  \BibitemOpen
  \bibfield  {author} {\bibinfo {author} {\bibfnamefont {V.~A.}\ \bibnamefont
  {Dzuba}}, \bibinfo {author} {\bibfnamefont {V.~V.}\ \bibnamefont {Flambaum}},
  \ and\ \bibinfo {author} {\bibfnamefont {Y.~V.}\ \bibnamefont {Stadnik}},\
  }\href {\doibase 10.1103/PhysRevLett.119.223201} {\bibfield  {journal}
  {\bibinfo  {journal} {Phys. Rev. Lett.}\ }\textbf {\bibinfo {volume} {119}},\
  \bibinfo {pages} {223201} (\bibinfo {year} {2017})}\BibitemShut {NoStop}%
\bibitem [{\citenamefont {Stadnik}\ and\ \citenamefont
  {Flambaum}(2014)}]{Stadnik2014}%
  \BibitemOpen
  \bibfield  {author} {\bibinfo {author} {\bibfnamefont {Y.~V.}\ \bibnamefont
  {Stadnik}}\ and\ \bibinfo {author} {\bibfnamefont {V.~V.}\ \bibnamefont
  {Flambaum}},\ }\href {\doibase 10.1103/PhysRevD.89.043522} {\bibfield
  {journal} {\bibinfo  {journal} {Phys. Rev. D}\ }\textbf {\bibinfo {volume}
  {89}},\ \bibinfo {pages} {043522} (\bibinfo {year} {2014})}\BibitemShut
  {NoStop}%
\bibitem [{\citenamefont {Kozlov}(1985)}]{Kozlov1986}%
  \BibitemOpen
  \bibfield  {author} {\bibinfo {author} {\bibfnamefont {M.~G.}\ \bibnamefont
  {Kozlov}},\ }\href@noop {} {\bibfield  {journal} {\bibinfo  {journal} {Sov.
  Phys. JETP}\ }\textbf {\bibinfo {volume} {62}},\ \bibinfo {pages} {1114}
  (\bibinfo {year} {1985})},\ \bibinfo {note} {[Zh. Eksp. Teor.
  Fiz.89,1933(1985)]}\BibitemShut {NoStop}%
\bibitem [{\citenamefont {Flambaum}\ and\ \citenamefont
  {Khriplovich}(1985)}]{Flambaum1985}%
  \BibitemOpen
  \bibfield  {author} {\bibinfo {author} {\bibfnamefont {V.}~\bibnamefont
  {Flambaum}}\ and\ \bibinfo {author} {\bibfnamefont {I.}~\bibnamefont
  {Khriplovich}},\ }\href@noop {} {\bibfield  {journal} {\bibinfo  {journal}
  {Phys. Lett. A}\ }\textbf {\bibinfo {volume} {110}},\ \bibinfo {pages} {121}
  (\bibinfo {year} {1985})}\BibitemShut {NoStop}%
\bibitem [{\citenamefont {DeMille}\ \emph
  {et~al.}(2008{\natexlab{a}})\citenamefont {DeMille}, \citenamefont {Cahn},
  \citenamefont {Murphree}, \citenamefont {Rahmlow},\ and\ \citenamefont
  {Kozlov}}]{DeMille2008b}%
  \BibitemOpen
  \bibfield  {author} {\bibinfo {author} {\bibfnamefont {D.}~\bibnamefont
  {DeMille}}, \bibinfo {author} {\bibfnamefont {S.~B.}\ \bibnamefont {Cahn}},
  \bibinfo {author} {\bibfnamefont {D.}~\bibnamefont {Murphree}}, \bibinfo
  {author} {\bibfnamefont {D.~A.}\ \bibnamefont {Rahmlow}}, \ and\ \bibinfo
  {author} {\bibfnamefont {M.~G.}\ \bibnamefont {Kozlov}},\ }\href {\doibase
  10.1103/PhysRevLett.100.023003} {\bibfield  {journal} {\bibinfo  {journal}
  {Phys. Rev. Lett.}\ }\textbf {\bibinfo {volume} {100}},\ \bibinfo {pages}
  {023003} (\bibinfo {year} {2008}{\natexlab{a}})}\BibitemShut {NoStop}%
\bibitem [{\citenamefont {Altunta\ifmmode~\mbox{\c{s}}\else \c{s}\fi{}}\ \emph
  {et~al.}(2018{\natexlab{a}})\citenamefont {Altunta\ifmmode~\mbox{\c{s}}\else
  \c{s}\fi{}}, \citenamefont {Ammon}, \citenamefont {Cahn},\ and\ \citenamefont
  {DeMille}}]{Altuntas2018}%
  \BibitemOpen
  \bibfield  {author} {\bibinfo {author} {\bibfnamefont {E.}~\bibnamefont
  {Altunta\ifmmode~\mbox{\c{s}}\else \c{s}\fi{}}}, \bibinfo {author}
  {\bibfnamefont {J.}~\bibnamefont {Ammon}}, \bibinfo {author} {\bibfnamefont
  {S.~B.}\ \bibnamefont {Cahn}}, \ and\ \bibinfo {author} {\bibfnamefont
  {D.}~\bibnamefont {DeMille}},\ }\href {\doibase
  10.1103/PhysRevLett.120.142501} {\bibfield  {journal} {\bibinfo  {journal}
  {Phys. Rev. Lett.}\ }\textbf {\bibinfo {volume} {120}},\ \bibinfo {pages}
  {142501} (\bibinfo {year} {2018}{\natexlab{a}})}\BibitemShut {NoStop}%
\bibitem [{\citenamefont {Altunta\ifmmode~\mbox{\c{s}}\else \c{s}\fi{}}\ \emph
  {et~al.}(2018{\natexlab{b}})\citenamefont {Altunta\ifmmode~\mbox{\c{s}}\else
  \c{s}\fi{}}, \citenamefont {Ammon}, \citenamefont {Cahn},\ and\ \citenamefont
  {DeMille}}]{Altuntas2018b}%
  \BibitemOpen
  \bibfield  {author} {\bibinfo {author} {\bibfnamefont {E.}~\bibnamefont
  {Altunta\ifmmode~\mbox{\c{s}}\else \c{s}\fi{}}}, \bibinfo {author}
  {\bibfnamefont {J.}~\bibnamefont {Ammon}}, \bibinfo {author} {\bibfnamefont
  {S.~B.}\ \bibnamefont {Cahn}}, \ and\ \bibinfo {author} {\bibfnamefont
  {D.}~\bibnamefont {DeMille}},\ }\href {\doibase 10.1103/PhysRevA.97.042101}
  {\bibfield  {journal} {\bibinfo  {journal} {Phys. Rev. A}\ }\textbf {\bibinfo
  {volume} {97}},\ \bibinfo {pages} {042101} (\bibinfo {year}
  {2018}{\natexlab{b}})}\BibitemShut {NoStop}%
\bibitem [{\citenamefont {Maris}\ \emph {et~al.}(2011)\citenamefont {Maris},
  \citenamefont {Vary}, \citenamefont {Navr\'atil}, \citenamefont {Ormand},
  \citenamefont {Nam},\ and\ \citenamefont {Dean}}]{Maris2011}%
  \BibitemOpen
  \bibfield  {author} {\bibinfo {author} {\bibfnamefont {P.}~\bibnamefont
  {Maris}}, \bibinfo {author} {\bibfnamefont {J.~P.}\ \bibnamefont {Vary}},
  \bibinfo {author} {\bibfnamefont {P.}~\bibnamefont {Navr\'atil}}, \bibinfo
  {author} {\bibfnamefont {W.~E.}\ \bibnamefont {Ormand}}, \bibinfo {author}
  {\bibfnamefont {H.}~\bibnamefont {Nam}}, \ and\ \bibinfo {author}
  {\bibfnamefont {D.~J.}\ \bibnamefont {Dean}},\ }\href {\doibase
  10.1103/PhysRevLett.106.202502} {\bibfield  {journal} {\bibinfo  {journal}
  {Phys. Rev. Lett.}\ }\textbf {\bibinfo {volume} {106}},\ \bibinfo {pages}
  {202502} (\bibinfo {year} {2011})}\BibitemShut {NoStop}%
\bibitem [{\citenamefont {Anderegg}\ \emph {et~al.}(2018)\citenamefont
  {Anderegg}, \citenamefont {Augenbraun}, \citenamefont {Bao}, \citenamefont
  {Burchesky}, \citenamefont {Cheuk}, \citenamefont {Ketterle},\ and\
  \citenamefont {Doyle}}]{Anderegg2018}%
  \BibitemOpen
  \bibfield  {author} {\bibinfo {author} {\bibfnamefont {L.}~\bibnamefont
  {Anderegg}}, \bibinfo {author} {\bibfnamefont {B.~L.}\ \bibnamefont
  {Augenbraun}}, \bibinfo {author} {\bibfnamefont {Y.}~\bibnamefont {Bao}},
  \bibinfo {author} {\bibfnamefont {S.}~\bibnamefont {Burchesky}}, \bibinfo
  {author} {\bibfnamefont {L.~W.}\ \bibnamefont {Cheuk}}, \bibinfo {author}
  {\bibfnamefont {W.}~\bibnamefont {Ketterle}}, \ and\ \bibinfo {author}
  {\bibfnamefont {J.~M.}\ \bibnamefont {Doyle}},\ }\href {\doibase
  10.1038/s41567-018-0191-z} {\bibfield  {journal} {\bibinfo  {journal} {Nature
  Physics}\ } (\bibinfo {year} {2018}),\ 10.1038/s41567-018-0191-z}\BibitemShut
  {NoStop}%
\bibitem [{\citenamefont {Prehn}\ \emph {et~al.}(2016)\citenamefont {Prehn},
  \citenamefont {Ibr\"ugger}, \citenamefont {Gl\"ockner}, \citenamefont
  {Rempe},\ and\ \citenamefont {Zeppenfeld}}]{Prehn2016}%
  \BibitemOpen
  \bibfield  {author} {\bibinfo {author} {\bibfnamefont {A.}~\bibnamefont
  {Prehn}}, \bibinfo {author} {\bibfnamefont {M.}~\bibnamefont {Ibr\"ugger}},
  \bibinfo {author} {\bibfnamefont {R.}~\bibnamefont {Gl\"ockner}}, \bibinfo
  {author} {\bibfnamefont {G.}~\bibnamefont {Rempe}}, \ and\ \bibinfo {author}
  {\bibfnamefont {M.}~\bibnamefont {Zeppenfeld}},\ }\href {\doibase
  10.1103/PhysRevLett.116.063005} {\bibfield  {journal} {\bibinfo  {journal}
  {Phys. Rev. Lett.}\ }\textbf {\bibinfo {volume} {116}},\ \bibinfo {pages}
  {063005} (\bibinfo {year} {2016})}\BibitemShut {NoStop}%
\bibitem [{\citenamefont {Kozyryev}\ \emph
  {et~al.}(2016{\natexlab{a}})\citenamefont {Kozyryev}, \citenamefont {Baum},
  \citenamefont {Matsuda}, \citenamefont {Hemmerling},\ and\ \citenamefont
  {Doyle}}]{Kozyryev2016}%
  \BibitemOpen
  \bibfield  {author} {\bibinfo {author} {\bibfnamefont {I.}~\bibnamefont
  {Kozyryev}}, \bibinfo {author} {\bibfnamefont {L.}~\bibnamefont {Baum}},
  \bibinfo {author} {\bibfnamefont {K.}~\bibnamefont {Matsuda}}, \bibinfo
  {author} {\bibfnamefont {B.}~\bibnamefont {Hemmerling}}, \ and\ \bibinfo
  {author} {\bibfnamefont {J.~M.}\ \bibnamefont {Doyle}},\ }\href
  {http://stacks.iop.org/0953-4075/49/i=13/a=134002} {\bibfield  {journal}
  {\bibinfo  {journal} {Journal of Physics B: Atomic, Molecular and Optical
  Physics}\ }\textbf {\bibinfo {volume} {49}},\ \bibinfo {pages} {134002}
  (\bibinfo {year} {2016}{\natexlab{a}})}\BibitemShut {NoStop}%
\bibitem [{\citenamefont {Kozyryev}\ \emph
  {et~al.}(2016{\natexlab{b}})\citenamefont {Kozyryev}, \citenamefont {Baum},
  \citenamefont {Matsuda},\ and\ \citenamefont {Doyle}}]{Kozyryev2016b}%
  \BibitemOpen
  \bibfield  {author} {\bibinfo {author} {\bibfnamefont {I.}~\bibnamefont
  {Kozyryev}}, \bibinfo {author} {\bibfnamefont {L.}~\bibnamefont {Baum}},
  \bibinfo {author} {\bibfnamefont {K.}~\bibnamefont {Matsuda}}, \ and\
  \bibinfo {author} {\bibfnamefont {J.~M.}\ \bibnamefont {Doyle}},\ }\href
  {\doibase 10.1002/cphc.201601051} {\bibfield  {journal} {\bibinfo  {journal}
  {ChemPhysChem}\ }\textbf {\bibinfo {volume} {17}},\ \bibinfo {pages} {3641}
  (\bibinfo {year} {2016}{\natexlab{b}})},\ \Eprint
  {http://arxiv.org/abs/https://onlinelibrary.wiley.com/doi/pdf/10.1002/cphc.201601051}
  {https://onlinelibrary.wiley.com/doi/pdf/10.1002/cphc.201601051} \BibitemShut
  {NoStop}%
\bibitem [{\citenamefont {Kozyryev}\ \emph {et~al.}(2017)\citenamefont
  {Kozyryev}, \citenamefont {Baum}, \citenamefont {Matsuda}, \citenamefont
  {Augenbraun}, \citenamefont {Anderegg}, \citenamefont {Sedlack},\ and\
  \citenamefont {Doyle}}]{Kozyryev2017}%
  \BibitemOpen
  \bibfield  {author} {\bibinfo {author} {\bibfnamefont {I.}~\bibnamefont
  {Kozyryev}}, \bibinfo {author} {\bibfnamefont {L.}~\bibnamefont {Baum}},
  \bibinfo {author} {\bibfnamefont {K.}~\bibnamefont {Matsuda}}, \bibinfo
  {author} {\bibfnamefont {B.~L.}\ \bibnamefont {Augenbraun}}, \bibinfo
  {author} {\bibfnamefont {L.}~\bibnamefont {Anderegg}}, \bibinfo {author}
  {\bibfnamefont {A.~P.}\ \bibnamefont {Sedlack}}, \ and\ \bibinfo {author}
  {\bibfnamefont {J.~M.}\ \bibnamefont {Doyle}},\ }\href {\doibase
  10.1103/PhysRevLett.118.173201} {\bibfield  {journal} {\bibinfo  {journal}
  {Phys. Rev. Lett.}\ }\textbf {\bibinfo {volume} {118}},\ \bibinfo {pages}
  {173201} (\bibinfo {year} {2017})}\BibitemShut {NoStop}%
\bibitem [{\citenamefont {Isaev}\ \emph {et~al.}(2017)\citenamefont {Isaev},
  \citenamefont {Zaitsevskii},\ and\ \citenamefont {Eliav}}]{Isaev2017}%
  \BibitemOpen
  \bibfield  {author} {\bibinfo {author} {\bibfnamefont {T.~A.}\ \bibnamefont
  {Isaev}}, \bibinfo {author} {\bibfnamefont {A.~V.}\ \bibnamefont
  {Zaitsevskii}}, \ and\ \bibinfo {author} {\bibfnamefont {E.}~\bibnamefont
  {Eliav}},\ }\href {http://stacks.iop.org/0953-4075/50/i=22/a=225101}
  {\bibfield  {journal} {\bibinfo  {journal} {Journal of Physics B: Atomic,
  Molecular and Optical Physics}\ }\textbf {\bibinfo {volume} {50}},\ \bibinfo
  {pages} {225101} (\bibinfo {year} {2017})}\BibitemShut {NoStop}%
\bibitem [{\citenamefont {Kozyryev}\ \emph {et~al.}(2018)\citenamefont
  {Kozyryev}, \citenamefont {Lasner},\ and\ \citenamefont
  {Doyle}}]{Kozyryev2018}%
  \BibitemOpen
  \bibfield  {author} {\bibinfo {author} {\bibfnamefont {I.}~\bibnamefont
  {Kozyryev}}, \bibinfo {author} {\bibfnamefont {Z.}~\bibnamefont {Lasner}}, \
  and\ \bibinfo {author} {\bibfnamefont {J.~M.}\ \bibnamefont {Doyle}},\
  }\href@noop {} {\bibfield  {journal} {\bibinfo  {journal} {ArXiv e-prints}\ }
  (\bibinfo {year} {2018})},\ \Eprint {http://arxiv.org/abs/1805.08185}
  {arXiv:1805.08185 [physics.atom-ph]} \BibitemShut {NoStop}%
\bibitem [{\citenamefont {Kozyryev}\ and\ \citenamefont
  {Hutzler}(2017)}]{Kozyryev2017b}%
  \BibitemOpen
  \bibfield  {author} {\bibinfo {author} {\bibfnamefont {I.}~\bibnamefont
  {Kozyryev}}\ and\ \bibinfo {author} {\bibfnamefont {N.~R.}\ \bibnamefont
  {Hutzler}},\ }\href {\doibase 10.1103/PhysRevLett.119.133002} {\bibfield
  {journal} {\bibinfo  {journal} {Phys. Rev. Lett.}\ }\textbf {\bibinfo
  {volume} {119}},\ \bibinfo {pages} {133002} (\bibinfo {year}
  {2017})}\BibitemShut {NoStop}%
\bibitem [{\citenamefont {Kozlov}(2013)}]{Kozlov2013}%
  \BibitemOpen
  \bibfield  {author} {\bibinfo {author} {\bibfnamefont {M.~G.}\ \bibnamefont
  {Kozlov}},\ }\href {\doibase 10.1103/PhysRevA.87.032104} {\bibfield
  {journal} {\bibinfo  {journal} {Phys. Rev. A}\ }\textbf {\bibinfo {volume}
  {87}},\ \bibinfo {pages} {032104} (\bibinfo {year} {2013})}\BibitemShut
  {NoStop}%
\bibitem [{\citenamefont {Alexander~Prehn}(2018)}]{Prehn2018}%
  \BibitemOpen
  \bibfield  {author} {\bibinfo {author} {\bibfnamefont {G.~R. M.~Z.}\
  \bibnamefont {Alexander~Prehn}, \bibfnamefont {Martin~Ibrugger}},\
  }\href@noop {} {\bibfield  {journal} {\bibinfo  {journal} {ArXiv e-prints}\ }
  (\bibinfo {year} {2018})},\ \Eprint {http://arxiv.org/abs/1807.06618}
  {arXiv:1807.06618 [physics.atom-ph]} \BibitemShut {NoStop}%
\bibitem [{\citenamefont {Park}\ \emph {et~al.}(2017)\citenamefont {Park},
  \citenamefont {Yan}, \citenamefont {Loh}, \citenamefont {Will},\ and\
  \citenamefont {Zwierlein}}]{Park2017}%
  \BibitemOpen
  \bibfield  {author} {\bibinfo {author} {\bibfnamefont {J.~W.}\ \bibnamefont
  {Park}}, \bibinfo {author} {\bibfnamefont {Z.~Z.}\ \bibnamefont {Yan}},
  \bibinfo {author} {\bibfnamefont {H.}~\bibnamefont {Loh}}, \bibinfo {author}
  {\bibfnamefont {S.~A.}\ \bibnamefont {Will}}, \ and\ \bibinfo {author}
  {\bibfnamefont {M.~W.}\ \bibnamefont {Zwierlein}},\ }\href {\doibase
  10.1126/science.aal5066} {\bibfield  {journal} {\bibinfo  {journal}
  {Science}\ }\textbf {\bibinfo {volume} {357}},\ \bibinfo {pages} {372}
  (\bibinfo {year} {2017})},\ \Eprint
  {http://arxiv.org/abs/http://science.sciencemag.org/content/357/6349/372.full.pdf}
  {http://science.sciencemag.org/content/357/6349/372.full.pdf} \BibitemShut
  {NoStop}%
\bibitem [{\citenamefont {Cheuk}\ \emph {et~al.}(2018)\citenamefont {Cheuk},
  \citenamefont {Anderegg}, \citenamefont {Augenbraun}, \citenamefont {Bao},
  \citenamefont {Burchesky}, \citenamefont {Ketterle},\ and\ \citenamefont
  {Doyle}}]{Cheuk2018}%
  \BibitemOpen
  \bibfield  {author} {\bibinfo {author} {\bibfnamefont {L.~W.}\ \bibnamefont
  {Cheuk}}, \bibinfo {author} {\bibfnamefont {L.}~\bibnamefont {Anderegg}},
  \bibinfo {author} {\bibfnamefont {B.~L.}\ \bibnamefont {Augenbraun}},
  \bibinfo {author} {\bibfnamefont {Y.}~\bibnamefont {Bao}}, \bibinfo {author}
  {\bibfnamefont {S.}~\bibnamefont {Burchesky}}, \bibinfo {author}
  {\bibfnamefont {W.}~\bibnamefont {Ketterle}}, \ and\ \bibinfo {author}
  {\bibfnamefont {J.~M.}\ \bibnamefont {Doyle}},\ }\href {\doibase
  10.1103/PhysRevLett.121.083201} {\bibfield  {journal} {\bibinfo  {journal}
  {Phys. Rev. Lett.}\ }\textbf {\bibinfo {volume} {121}},\ \bibinfo {pages}
  {083201} (\bibinfo {year} {2018})}\BibitemShut {NoStop}%
\bibitem [{\citenamefont {Cahn}\ \emph {et~al.}(2014)\citenamefont {Cahn},
  \citenamefont {Ammon}, \citenamefont {Kirilov}, \citenamefont {Gurevich},
  \citenamefont {Murphree}, \citenamefont {Paolino}, \citenamefont {Rahmlow},
  \citenamefont {Kozlov},\ and\ \citenamefont {DeMille}}]{Cahn2014}%
  \BibitemOpen
  \bibfield  {author} {\bibinfo {author} {\bibfnamefont {S.~B.}\ \bibnamefont
  {Cahn}}, \bibinfo {author} {\bibfnamefont {J.}~\bibnamefont {Ammon}},
  \bibinfo {author} {\bibfnamefont {E.}~\bibnamefont {Kirilov}}, \bibinfo
  {author} {\bibfnamefont {Y.~V.}\ \bibnamefont {Gurevich}}, \bibinfo {author}
  {\bibfnamefont {D.}~\bibnamefont {Murphree}}, \bibinfo {author}
  {\bibfnamefont {R.}~\bibnamefont {Paolino}}, \bibinfo {author} {\bibfnamefont
  {D.~A.}\ \bibnamefont {Rahmlow}}, \bibinfo {author} {\bibfnamefont {M.~G.}\
  \bibnamefont {Kozlov}}, \ and\ \bibinfo {author} {\bibfnamefont
  {D.}~\bibnamefont {DeMille}},\ }\href {\doibase
  10.1103/PhysRevLett.112.163002} {\bibfield  {journal} {\bibinfo  {journal}
  {Phys. Rev. Lett.}\ }\textbf {\bibinfo {volume} {112}},\ \bibinfo {pages}
  {163002} (\bibinfo {year} {2014})}\BibitemShut {NoStop}%
\bibitem [{\citenamefont {Hirota}(1985)}]{Hirota1985}%
  \BibitemOpen
  \bibfield  {author} {\bibinfo {author} {\bibfnamefont {E.}~\bibnamefont
  {Hirota}},\ }\href@noop {} {\emph {\bibinfo {title} {High-Resolution
  Spectroscopy of Transient Molecules}}}\ (\bibinfo  {publisher}
  {Springer-Verlag},\ \bibinfo {address} {Berlin},\ \bibinfo {year}
  {1985})\BibitemShut {NoStop}%
\bibitem [{\citenamefont {Brown}\ and\ \citenamefont
  {Carrington}(2003)}]{Brown2003}%
  \BibitemOpen
  \bibfield  {author} {\bibinfo {author} {\bibfnamefont {J.~M.}\ \bibnamefont
  {Brown}}\ and\ \bibinfo {author} {\bibfnamefont {A.}~\bibnamefont
  {Carrington}},\ }\href@noop {} {\emph {\bibinfo {title} {Rotational
  spectroscopy of diatomic molecules}}}\ (\bibinfo  {publisher} {Cambridge
  Univ. Press},\ \bibinfo {year} {2003})\BibitemShut {NoStop}%
\bibitem [{\citenamefont {Flambaum}\ and\ \citenamefont
  {Khriplovich}(1980)}]{Flambaum1980}%
  \BibitemOpen
  \bibfield  {author} {\bibinfo {author} {\bibfnamefont {V.}~\bibnamefont
  {Flambaum}}\ and\ \bibinfo {author} {\bibfnamefont {I.}~\bibnamefont
  {Khriplovich}},\ }\href@noop {} {\bibfield  {journal} {\bibinfo  {journal}
  {Sov. Phys. JETP}\ }\textbf {\bibinfo {volume} {52}},\ \bibinfo {pages} {835}
  (\bibinfo {year} {1980})}\BibitemShut {NoStop}%
\bibitem [{\citenamefont {Sheng}\ \emph {et~al.}(2010)\citenamefont {Sheng},
  \citenamefont {Orozco},\ and\ \citenamefont {Gomez}}]{Sheng2010}%
  \BibitemOpen
  \bibfield  {author} {\bibinfo {author} {\bibfnamefont {D.}~\bibnamefont
  {Sheng}}, \bibinfo {author} {\bibfnamefont {L.~A.}\ \bibnamefont {Orozco}}, \
  and\ \bibinfo {author} {\bibfnamefont {E.}~\bibnamefont {Gomez}},\ }\href
  {http://stacks.iop.org/0953-4075/43/i=7/a=074004} {\bibfield  {journal}
  {\bibinfo  {journal} {Journal of Physics B: Atomic, Molecular and Optical
  Physics}\ }\textbf {\bibinfo {volume} {43}},\ \bibinfo {pages} {074004}
  (\bibinfo {year} {2010})}\BibitemShut {NoStop}%
\bibitem [{\citenamefont {Borschevsky}\ \emph {et~al.}(2013)\citenamefont
  {Borschevsky}, \citenamefont {Ilia\ifmmode~\check{s}\else \v{s}\fi{}},
  \citenamefont {Dzuba}, \citenamefont {Flambaum},\ and\ \citenamefont
  {Schwerdtfeger}}]{Borschevsky2013}%
  \BibitemOpen
  \bibfield  {author} {\bibinfo {author} {\bibfnamefont {A.}~\bibnamefont
  {Borschevsky}}, \bibinfo {author} {\bibfnamefont {M.}~\bibnamefont
  {Ilia\ifmmode~\check{s}\else \v{s}\fi{}}}, \bibinfo {author} {\bibfnamefont
  {V.~A.}\ \bibnamefont {Dzuba}}, \bibinfo {author} {\bibfnamefont {V.~V.}\
  \bibnamefont {Flambaum}}, \ and\ \bibinfo {author} {\bibfnamefont
  {P.}~\bibnamefont {Schwerdtfeger}},\ }\href {\doibase
  10.1103/PhysRevA.88.022125} {\bibfield  {journal} {\bibinfo  {journal} {Phys.
  Rev. A}\ }\textbf {\bibinfo {volume} {88}},\ \bibinfo {pages} {022125}
  (\bibinfo {year} {2013})}\BibitemShut {NoStop}%
\bibitem [{\citenamefont {Hao}\ \emph {et~al.}(2018)\citenamefont {Hao},
  \citenamefont {Ilias}, \citenamefont {Eliav}, \citenamefont {Schwerdtfeger},
  \citenamefont {Flambaum},\ and\ \citenamefont {Borschevsky}}]{Hao2018}%
  \BibitemOpen
  \bibfield  {author} {\bibinfo {author} {\bibfnamefont {Y.}~\bibnamefont
  {Hao}}, \bibinfo {author} {\bibfnamefont {M.}~\bibnamefont {Ilias}}, \bibinfo
  {author} {\bibfnamefont {E.}~\bibnamefont {Eliav}}, \bibinfo {author}
  {\bibfnamefont {P.}~\bibnamefont {Schwerdtfeger}}, \bibinfo {author}
  {\bibfnamefont {V.}~\bibnamefont {Flambaum}}, \ and\ \bibinfo {author}
  {\bibfnamefont {A.}~\bibnamefont {Borschevsky}},\ }\href@noop {} {\bibfield
  {journal} {\bibinfo  {journal} {ArXiv e-prints}\ } (\bibinfo {year}
  {2018})},\ \Eprint {http://arxiv.org/abs/1808.02771} {arXiv:1808.02771
  [physics.atom-ph]} \BibitemShut {NoStop}%
\bibitem [{\citenamefont {Khriplovich}(1991)}]{khriplovich1991}%
  \BibitemOpen
  \bibfield  {author} {\bibinfo {author} {\bibfnamefont {I.~B.}\ \bibnamefont
  {Khriplovich}},\ }\href
  {http://inis.iaea.org/search/search.aspx?orig_q=RN:25004519} {\emph {\bibinfo
  {title} {Parity nonconservation in atomic phenomena}}}\ (\bibinfo
  {publisher} {Gordon and Breach Science Publishers},\ \bibinfo {address}
  {United States},\ \bibinfo {year} {1991})\BibitemShut {NoStop}%
\bibitem [{\citenamefont {Di~Rosa}(2004)}]{DiRosa2004}%
  \BibitemOpen
  \bibfield  {author} {\bibinfo {author} {\bibfnamefont {M.}~\bibnamefont
  {Di~Rosa}},\ }\href {\doibase 10.1140/epjd/e2004-00167-2} {\bibfield
  {journal} {\bibinfo  {journal} {The European Physical Journal D - Atomic,
  Molecular, Optical and Plasma Physics}\ }\textbf {\bibinfo {volume} {31}},\
  \bibinfo {pages} {395} (\bibinfo {year} {2004})}\BibitemShut {NoStop}%
\bibitem [{\citenamefont {{Lane}}(2012)}]{Lane2012}%
  \BibitemOpen
  \bibfield  {author} {\bibinfo {author} {\bibfnamefont {I.~C.}\ \bibnamefont
  {{Lane}}},\ }\href {\doibase 10.1039/c2cp42709e} {\bibfield  {journal}
  {\bibinfo  {journal} {Physical Chemistry Chemical Physics (Incorporating
  Faraday Transactions)}\ }\textbf {\bibinfo {volume} {14}},\ \bibinfo {pages}
  {15078} (\bibinfo {year} {2012})}\BibitemShut {NoStop}%
\bibitem [{\citenamefont {Hendricks}\ \emph {et~al.}(2014)\citenamefont
  {Hendricks}, \citenamefont {Holland}, \citenamefont {Truppe}, \citenamefont
  {Sauer},\ and\ \citenamefont {Tarbutt}}]{Hendricks2014}%
  \BibitemOpen
  \bibfield  {author} {\bibinfo {author} {\bibfnamefont {R.}~\bibnamefont
  {Hendricks}}, \bibinfo {author} {\bibfnamefont {D.}~\bibnamefont {Holland}},
  \bibinfo {author} {\bibfnamefont {S.}~\bibnamefont {Truppe}}, \bibinfo
  {author} {\bibfnamefont {B.~E.}\ \bibnamefont {Sauer}}, \ and\ \bibinfo
  {author} {\bibfnamefont {M.}~\bibnamefont {Tarbutt}},\ }\href
  {http://www.frontiersin.org/physical_chemistry_and_chemical_physics/10.3389/fphy.2014.00051/abstract}
  {\bibfield  {journal} {\bibinfo  {journal} {Frontiers in Physics}\ }\textbf
  {\bibinfo {volume} {2}} (\bibinfo {year} {2014})}\BibitemShut {NoStop}%
\bibitem [{\citenamefont {Lim}\ \emph {et~al.}(2018)\citenamefont {Lim},
  \citenamefont {Almond}, \citenamefont {Trigatzis}, \citenamefont {Devlin},
  \citenamefont {Fitch}, \citenamefont {Sauer}, \citenamefont {Tarbutt},\ and\
  \citenamefont {Hinds}}]{Lim2018}%
  \BibitemOpen
  \bibfield  {author} {\bibinfo {author} {\bibfnamefont {J.}~\bibnamefont
  {Lim}}, \bibinfo {author} {\bibfnamefont {J.~R.}\ \bibnamefont {Almond}},
  \bibinfo {author} {\bibfnamefont {M.~A.}\ \bibnamefont {Trigatzis}}, \bibinfo
  {author} {\bibfnamefont {J.~A.}\ \bibnamefont {Devlin}}, \bibinfo {author}
  {\bibfnamefont {N.~J.}\ \bibnamefont {Fitch}}, \bibinfo {author}
  {\bibfnamefont {B.~E.}\ \bibnamefont {Sauer}}, \bibinfo {author}
  {\bibfnamefont {M.~R.}\ \bibnamefont {Tarbutt}}, \ and\ \bibinfo {author}
  {\bibfnamefont {E.~A.}\ \bibnamefont {Hinds}},\ }\href {\doibase
  10.1103/PhysRevLett.120.123201} {\bibfield  {journal} {\bibinfo  {journal}
  {Phys. Rev. Lett.}\ }\textbf {\bibinfo {volume} {120}},\ \bibinfo {pages}
  {123201} (\bibinfo {year} {2018})}\BibitemShut {NoStop}%
\bibitem [{\citenamefont {Hunter}\ \emph {et~al.}(2012)\citenamefont {Hunter},
  \citenamefont {Peck}, \citenamefont {Greenspon}, \citenamefont {Alam},\ and\
  \citenamefont {DeMille}}]{Hunter2012}%
  \BibitemOpen
  \bibfield  {author} {\bibinfo {author} {\bibfnamefont {L.~R.}\ \bibnamefont
  {Hunter}}, \bibinfo {author} {\bibfnamefont {S.~K.}\ \bibnamefont {Peck}},
  \bibinfo {author} {\bibfnamefont {A.~S.}\ \bibnamefont {Greenspon}}, \bibinfo
  {author} {\bibfnamefont {S.~S.}\ \bibnamefont {Alam}}, \ and\ \bibinfo
  {author} {\bibfnamefont {D.}~\bibnamefont {DeMille}},\ }\href {\doibase
  10.1103/PhysRevA.85.012511} {\bibfield  {journal} {\bibinfo  {journal} {Phys.
  Rev. A}\ }\textbf {\bibinfo {volume} {85}},\ \bibinfo {pages} {012511}
  (\bibinfo {year} {2012})}\BibitemShut {NoStop}%
\bibitem [{\citenamefont {Norrgard}\ \emph {et~al.}(2017)\citenamefont
  {Norrgard}, \citenamefont {Edwards}, \citenamefont {McCarron}, \citenamefont
  {Steinecker}, \citenamefont {DeMille}, \citenamefont {Alam}, \citenamefont
  {Peck}, \citenamefont {Wadia},\ and\ \citenamefont {Hunter}}]{Norrgard2017}%
  \BibitemOpen
  \bibfield  {author} {\bibinfo {author} {\bibfnamefont {E.~B.}\ \bibnamefont
  {Norrgard}}, \bibinfo {author} {\bibfnamefont {E.~R.}\ \bibnamefont
  {Edwards}}, \bibinfo {author} {\bibfnamefont {D.~J.}\ \bibnamefont
  {McCarron}}, \bibinfo {author} {\bibfnamefont {M.~H.}\ \bibnamefont
  {Steinecker}}, \bibinfo {author} {\bibfnamefont {D.}~\bibnamefont {DeMille}},
  \bibinfo {author} {\bibfnamefont {S.~S.}\ \bibnamefont {Alam}}, \bibinfo
  {author} {\bibfnamefont {S.~K.}\ \bibnamefont {Peck}}, \bibinfo {author}
  {\bibfnamefont {N.~S.}\ \bibnamefont {Wadia}}, \ and\ \bibinfo {author}
  {\bibfnamefont {L.~R.}\ \bibnamefont {Hunter}},\ }\href {\doibase
  10.1103/PhysRevA.95.062506} {\bibfield  {journal} {\bibinfo  {journal} {Phys.
  Rev. A}\ }\textbf {\bibinfo {volume} {95}},\ \bibinfo {pages} {062506}
  (\bibinfo {year} {2017})}\BibitemShut {NoStop}%
\bibitem [{\citenamefont {Nakhate}\ \emph {et~al.}(2018)\citenamefont
  {Nakhate}, \citenamefont {Steimle}, \citenamefont {Pilgram},\ and\
  \citenamefont {Hutzler}}]{Nakhate2018}%
  \BibitemOpen
  \bibfield  {author} {\bibinfo {author} {\bibfnamefont {S.}~\bibnamefont
  {Nakhate}}, \bibinfo {author} {\bibfnamefont {T.~C.}\ \bibnamefont
  {Steimle}}, \bibinfo {author} {\bibfnamefont {N.~H.}\ \bibnamefont
  {Pilgram}}, \ and\ \bibinfo {author} {\bibfnamefont {N.~R.}\ \bibnamefont
  {Hutzler}},\ }\href@noop {} {\bibfield  {journal} {\bibinfo  {journal} {ArXiv
  e-prints}\ } (\bibinfo {year} {2018})},\ \Eprint
  {http://arxiv.org/abs/1810.02791} {arXiv:1810.02791 [physics.atom-ph]}
  \BibitemShut {NoStop}%
\bibitem [{\citenamefont {Maxwell}\ \emph {et~al.}(2005)\citenamefont
  {Maxwell}, \citenamefont {Brahms}, \citenamefont {deCarvalho}, \citenamefont
  {Glenn}, \citenamefont {Helton}, \citenamefont {Nguyen}, \citenamefont
  {Patterson}, \citenamefont {Petricka}, \citenamefont {DeMille},\ and\
  \citenamefont {Doyle}}]{Maxwell2005}%
  \BibitemOpen
  \bibfield  {author} {\bibinfo {author} {\bibfnamefont {S.~E.}\ \bibnamefont
  {Maxwell}}, \bibinfo {author} {\bibfnamefont {N.}~\bibnamefont {Brahms}},
  \bibinfo {author} {\bibfnamefont {R.}~\bibnamefont {deCarvalho}}, \bibinfo
  {author} {\bibfnamefont {D.~R.}\ \bibnamefont {Glenn}}, \bibinfo {author}
  {\bibfnamefont {J.~S.}\ \bibnamefont {Helton}}, \bibinfo {author}
  {\bibfnamefont {S.~V.}\ \bibnamefont {Nguyen}}, \bibinfo {author}
  {\bibfnamefont {D.}~\bibnamefont {Patterson}}, \bibinfo {author}
  {\bibfnamefont {J.}~\bibnamefont {Petricka}}, \bibinfo {author}
  {\bibfnamefont {D.}~\bibnamefont {DeMille}}, \ and\ \bibinfo {author}
  {\bibfnamefont {J.~M.}\ \bibnamefont {Doyle}},\ }\href {\doibase
  10.1103/PhysRevLett.95.173201} {\bibfield  {journal} {\bibinfo  {journal}
  {Phys. Rev. Lett.}\ }\textbf {\bibinfo {volume} {95}},\ \bibinfo {pages}
  {173201} (\bibinfo {year} {2005})}\BibitemShut {NoStop}%
\bibitem [{\citenamefont {Hutzler}\ \emph {et~al.}(2012)\citenamefont
  {Hutzler}, \citenamefont {Lu},\ and\ \citenamefont {Doyle}}]{Hutzler2012}%
  \BibitemOpen
  \bibfield  {author} {\bibinfo {author} {\bibfnamefont {N.~R.}\ \bibnamefont
  {Hutzler}}, \bibinfo {author} {\bibfnamefont {H.-I.}\ \bibnamefont {Lu}}, \
  and\ \bibinfo {author} {\bibfnamefont {J.~M.}\ \bibnamefont {Doyle}},\ }\href
  {\doibase 10.1021/cr200362u} {\bibfield  {journal} {\bibinfo  {journal}
  {Chemical Reviews}\ }\textbf {\bibinfo {volume} {112}},\ \bibinfo {pages}
  {4803} (\bibinfo {year} {2012})}\BibitemShut {NoStop}%
\bibitem [{\citenamefont {{Barry}}\ \emph {et~al.}(2012)\citenamefont
  {{Barry}}, \citenamefont {{Shuman}}, \citenamefont {{Norrgard}},\ and\
  \citenamefont {{DeMille}}}]{Barry2012}%
  \BibitemOpen
  \bibfield  {author} {\bibinfo {author} {\bibfnamefont {J.~F.}\ \bibnamefont
  {{Barry}}}, \bibinfo {author} {\bibfnamefont {E.~S.}\ \bibnamefont
  {{Shuman}}}, \bibinfo {author} {\bibfnamefont {E.~B.}\ \bibnamefont
  {{Norrgard}}}, \ and\ \bibinfo {author} {\bibfnamefont {D.}~\bibnamefont
  {{DeMille}}},\ }\href {\doibase 10.1103/PhysRevLett.108.103002} {\bibfield
  {journal} {\bibinfo  {journal} {Phys. Rev. Lett.}\ }\textbf {\bibinfo
  {volume} {108}},\ \bibinfo {eid} {103002} (\bibinfo {year}
  {2012})}\BibitemShut {NoStop}%
\bibitem [{\citenamefont {Barry}\ \emph {et~al.}(2014)\citenamefont {Barry},
  \citenamefont {McCarron}, \citenamefont {Norrgard}, \citenamefont
  {Steinecker},\ and\ \citenamefont {DeMille}}]{Barry2014}%
  \BibitemOpen
  \bibfield  {author} {\bibinfo {author} {\bibfnamefont {J.~F.}\ \bibnamefont
  {Barry}}, \bibinfo {author} {\bibfnamefont {D.~J.}\ \bibnamefont {McCarron}},
  \bibinfo {author} {\bibfnamefont {E.~N.}\ \bibnamefont {Norrgard}}, \bibinfo
  {author} {\bibfnamefont {M.~H.}\ \bibnamefont {Steinecker}}, \ and\ \bibinfo
  {author} {\bibfnamefont {D.}~\bibnamefont {DeMille}},\ }\href {\doibase
  10.1038/nature13634} {\bibfield  {journal} {\bibinfo  {journal} {Nature}\
  }\textbf {\bibinfo {volume} {512}},\ \bibinfo {pages} {286} (\bibinfo {year}
  {2014})}\BibitemShut {NoStop}%
\bibitem [{\citenamefont {Chotia}\ \emph {et~al.}(2012)\citenamefont {Chotia},
  \citenamefont {Neyenhuis}, \citenamefont {Moses}, \citenamefont {Yan},
  \citenamefont {Covey}, \citenamefont {Foss-Feig}, \citenamefont {Rey},
  \citenamefont {Jin},\ and\ \citenamefont {Ye}}]{Chotia2012}%
  \BibitemOpen
  \bibfield  {author} {\bibinfo {author} {\bibfnamefont {A.}~\bibnamefont
  {Chotia}}, \bibinfo {author} {\bibfnamefont {B.}~\bibnamefont {Neyenhuis}},
  \bibinfo {author} {\bibfnamefont {S.~A.}\ \bibnamefont {Moses}}, \bibinfo
  {author} {\bibfnamefont {B.}~\bibnamefont {Yan}}, \bibinfo {author}
  {\bibfnamefont {J.~P.}\ \bibnamefont {Covey}}, \bibinfo {author}
  {\bibfnamefont {M.}~\bibnamefont {Foss-Feig}}, \bibinfo {author}
  {\bibfnamefont {A.~M.}\ \bibnamefont {Rey}}, \bibinfo {author} {\bibfnamefont
  {D.~S.}\ \bibnamefont {Jin}}, \ and\ \bibinfo {author} {\bibfnamefont
  {J.}~\bibnamefont {Ye}},\ }\href {\doibase 10.1103/PhysRevLett.108.080405}
  {\bibfield  {journal} {\bibinfo  {journal} {Phys. Rev. Lett.}\ }\textbf
  {\bibinfo {volume} {108}},\ \bibinfo {pages} {080405} (\bibinfo {year}
  {2012})}\BibitemShut {NoStop}%
\bibitem [{\citenamefont {Panda}\ \emph {et~al.}(2016)\citenamefont {Panda},
  \citenamefont {O'Leary}, \citenamefont {West}, \citenamefont {Baron},
  \citenamefont {Hess}, \citenamefont {Hoffman}, \citenamefont {Kirilov},
  \citenamefont {Overstreet}, \citenamefont {West}, \citenamefont {DeMille},
  \citenamefont {Doyle},\ and\ \citenamefont {Gabrielse}}]{Panda2016}%
  \BibitemOpen
  \bibfield  {author} {\bibinfo {author} {\bibfnamefont {C.~D.}\ \bibnamefont
  {Panda}}, \bibinfo {author} {\bibfnamefont {B.~R.}\ \bibnamefont {O'Leary}},
  \bibinfo {author} {\bibfnamefont {A.~D.}\ \bibnamefont {West}}, \bibinfo
  {author} {\bibfnamefont {J.}~\bibnamefont {Baron}}, \bibinfo {author}
  {\bibfnamefont {P.~W.}\ \bibnamefont {Hess}}, \bibinfo {author}
  {\bibfnamefont {C.}~\bibnamefont {Hoffman}}, \bibinfo {author} {\bibfnamefont
  {E.}~\bibnamefont {Kirilov}}, \bibinfo {author} {\bibfnamefont {C.~B.}\
  \bibnamefont {Overstreet}}, \bibinfo {author} {\bibfnamefont {E.~P.}\
  \bibnamefont {West}}, \bibinfo {author} {\bibfnamefont {D.}~\bibnamefont
  {DeMille}}, \bibinfo {author} {\bibfnamefont {J.~M.}\ \bibnamefont {Doyle}},
  \ and\ \bibinfo {author} {\bibfnamefont {G.}~\bibnamefont {Gabrielse}},\
  }\href {\doibase 10.1103/PhysRevA.93.052110} {\bibfield  {journal} {\bibinfo
  {journal} {Phys. Rev. A}\ }\textbf {\bibinfo {volume} {93}},\ \bibinfo
  {pages} {052110} (\bibinfo {year} {2016})}\BibitemShut {NoStop}%
\bibitem [{\citenamefont {DeMille}\ \emph
  {et~al.}(2008{\natexlab{b}})\citenamefont {DeMille}, \citenamefont {Cahn},
  \citenamefont {Murphree}, \citenamefont {Rahmlow},\ and\ \citenamefont
  {Kozlov}}]{DeMille08b}%
  \BibitemOpen
  \bibfield  {author} {\bibinfo {author} {\bibfnamefont {D.}~\bibnamefont
  {DeMille}}, \bibinfo {author} {\bibfnamefont {S.~B.}\ \bibnamefont {Cahn}},
  \bibinfo {author} {\bibfnamefont {D.}~\bibnamefont {Murphree}}, \bibinfo
  {author} {\bibfnamefont {D.~A.}\ \bibnamefont {Rahmlow}}, \ and\ \bibinfo
  {author} {\bibfnamefont {M.~G.}\ \bibnamefont {Kozlov}},\ }\href@noop {}
  {\bibfield  {journal} {\bibinfo  {journal} {Phys. Rev. Lett.}\ }\textbf
  {\bibinfo {volume} {100}},\ \bibinfo {pages} {023003} (\bibinfo {year}
  {2008}{\natexlab{b}})}\BibitemShut {NoStop}%
\bibitem [{\citenamefont {Romalis}\ and\ \citenamefont
  {Fortson}(1999)}]{Romalis1999}%
  \BibitemOpen
  \bibfield  {author} {\bibinfo {author} {\bibfnamefont {M.~V.}\ \bibnamefont
  {Romalis}}\ and\ \bibinfo {author} {\bibfnamefont {E.~N.}\ \bibnamefont
  {Fortson}},\ }\href {\doibase 10.1103/PhysRevA.59.4547} {\bibfield  {journal}
  {\bibinfo  {journal} {Phys. Rev. A}\ }\textbf {\bibinfo {volume} {59}},\
  \bibinfo {pages} {4547} (\bibinfo {year} {1999})}\BibitemShut {NoStop}%
\bibitem [{\citenamefont {Kotochigova}\ and\ \citenamefont
  {DeMille}(2010)}]{Kotochigova2010}%
  \BibitemOpen
  \bibfield  {author} {\bibinfo {author} {\bibfnamefont {S.}~\bibnamefont
  {Kotochigova}}\ and\ \bibinfo {author} {\bibfnamefont {D.}~\bibnamefont
  {DeMille}},\ }\href {\doibase 10.1103/PhysRevA.82.063421} {\bibfield
  {journal} {\bibinfo  {journal} {Phys. Rev. A}\ }\textbf {\bibinfo {volume}
  {82}},\ \bibinfo {pages} {063421} (\bibinfo {year} {2010})}\BibitemShut
  {NoStop}%
\bibitem [{\citenamefont {Neyenhuis}\ \emph {et~al.}(2012)\citenamefont
  {Neyenhuis}, \citenamefont {Yan}, \citenamefont {Moses}, \citenamefont
  {Covey}, \citenamefont {Chotia}, \citenamefont {Petrov}, \citenamefont
  {Kotochigova}, \citenamefont {Ye},\ and\ \citenamefont
  {Jin}}]{Neyenhuis2012}%
  \BibitemOpen
  \bibfield  {author} {\bibinfo {author} {\bibfnamefont {B.}~\bibnamefont
  {Neyenhuis}}, \bibinfo {author} {\bibfnamefont {B.}~\bibnamefont {Yan}},
  \bibinfo {author} {\bibfnamefont {S.~A.}\ \bibnamefont {Moses}}, \bibinfo
  {author} {\bibfnamefont {J.~P.}\ \bibnamefont {Covey}}, \bibinfo {author}
  {\bibfnamefont {A.}~\bibnamefont {Chotia}}, \bibinfo {author} {\bibfnamefont
  {A.}~\bibnamefont {Petrov}}, \bibinfo {author} {\bibfnamefont
  {S.}~\bibnamefont {Kotochigova}}, \bibinfo {author} {\bibfnamefont
  {J.}~\bibnamefont {Ye}}, \ and\ \bibinfo {author} {\bibfnamefont {D.~S.}\
  \bibnamefont {Jin}},\ }\href {\doibase 10.1103/PhysRevLett.109.230403}
  {\bibfield  {journal} {\bibinfo  {journal} {Phys. Rev. Lett.}\ }\textbf
  {\bibinfo {volume} {109}},\ \bibinfo {pages} {230403} (\bibinfo {year}
  {2012})}\BibitemShut {NoStop}%
\bibitem [{\citenamefont {Kim}\ \emph {et~al.}(2013)\citenamefont {Kim},
  \citenamefont {Han},\ and\ \citenamefont {Cho}}]{Kim2013}%
  \BibitemOpen
  \bibfield  {author} {\bibinfo {author} {\bibfnamefont {H.}~\bibnamefont
  {Kim}}, \bibinfo {author} {\bibfnamefont {H.~S.}\ \bibnamefont {Han}}, \ and\
  \bibinfo {author} {\bibfnamefont {D.}~\bibnamefont {Cho}},\ }\href {\doibase
  10.1103/PhysRevLett.111.243004} {\bibfield  {journal} {\bibinfo  {journal}
  {Phys. Rev. Lett.}\ }\textbf {\bibinfo {volume} {111}},\ \bibinfo {pages}
  {243004} (\bibinfo {year} {2013})}\BibitemShut {NoStop}%
\bibitem [{\citenamefont {Rosenband}\ \emph {et~al.}(2018)\citenamefont
  {Rosenband}, \citenamefont {Grimes},\ and\ \citenamefont
  {Ni}}]{Rosenband2018}%
  \BibitemOpen
  \bibfield  {author} {\bibinfo {author} {\bibfnamefont {T.}~\bibnamefont
  {Rosenband}}, \bibinfo {author} {\bibfnamefont {D.~D.}\ \bibnamefont
  {Grimes}}, \ and\ \bibinfo {author} {\bibfnamefont {K.-K.}\ \bibnamefont
  {Ni}},\ }\href {\doibase 10.1364/OE.26.019821} {\bibfield  {journal}
  {\bibinfo  {journal} {Opt. Express}\ }\textbf {\bibinfo {volume} {26}},\
  \bibinfo {pages} {19821} (\bibinfo {year} {2018})}\BibitemShut {NoStop}%
\bibitem [{\citenamefont {O'Leary}(2016)}]{OlearyThesis}%
  \BibitemOpen
  \bibfield  {author} {\bibinfo {author} {\bibfnamefont {B.}~\bibnamefont
  {O'Leary}},\ }\emph {\bibinfo {title} {In search of the electron’s electric
  dipole moment in thorium monoxide: an improved upper limit, systematic error
  models, and apparatus upgrades}},\ \href {http://www.electronedm.org/} {Ph.D.
  thesis},\ \bibinfo  {school} {Yale University} (\bibinfo {year}
  {2016})\BibitemShut {NoStop}%
\bibitem [{\citenamefont {Norrgard}\ \emph {et~al.}(2016)\citenamefont
  {Norrgard}, \citenamefont {McCarron}, \citenamefont {Steinecker},
  \citenamefont {Tarbutt},\ and\ \citenamefont {DeMille}}]{Norrgard2016}%
  \BibitemOpen
  \bibfield  {author} {\bibinfo {author} {\bibfnamefont {E.~B.}\ \bibnamefont
  {Norrgard}}, \bibinfo {author} {\bibfnamefont {D.~J.}\ \bibnamefont
  {McCarron}}, \bibinfo {author} {\bibfnamefont {M.~H.}\ \bibnamefont
  {Steinecker}}, \bibinfo {author} {\bibfnamefont {M.~R.}\ \bibnamefont
  {Tarbutt}}, \ and\ \bibinfo {author} {\bibfnamefont {D.}~\bibnamefont
  {DeMille}},\ }\href {\doibase 10.1103/PhysRevLett.116.063004} {\bibfield
  {journal} {\bibinfo  {journal} {Phys. Rev. Lett.}\ }\textbf {\bibinfo
  {volume} {116}},\ \bibinfo {pages} {063004} (\bibinfo {year}
  {2016})}\BibitemShut {NoStop}%
\bibitem [{\citenamefont {Williams}\ \emph {et~al.}(2018)\citenamefont
  {Williams}, \citenamefont {Caldwell}, \citenamefont {Fitch}, \citenamefont
  {Truppe}, \citenamefont {Rodewald}, \citenamefont {Hinds}, \citenamefont
  {Sauer},\ and\ \citenamefont {Tarbutt}}]{Williams2018}%
  \BibitemOpen
  \bibfield  {author} {\bibinfo {author} {\bibfnamefont {H.~J.}\ \bibnamefont
  {Williams}}, \bibinfo {author} {\bibfnamefont {L.}~\bibnamefont {Caldwell}},
  \bibinfo {author} {\bibfnamefont {N.~J.}\ \bibnamefont {Fitch}}, \bibinfo
  {author} {\bibfnamefont {S.}~\bibnamefont {Truppe}}, \bibinfo {author}
  {\bibfnamefont {J.}~\bibnamefont {Rodewald}}, \bibinfo {author}
  {\bibfnamefont {E.~A.}\ \bibnamefont {Hinds}}, \bibinfo {author}
  {\bibfnamefont {B.~E.}\ \bibnamefont {Sauer}}, \ and\ \bibinfo {author}
  {\bibfnamefont {M.~R.}\ \bibnamefont {Tarbutt}},\ }\href {\doibase
  10.1103/PhysRevLett.120.163201} {\bibfield  {journal} {\bibinfo  {journal}
  {Phys. Rev. Lett.}\ }\textbf {\bibinfo {volume} {120}},\ \bibinfo {pages}
  {163201} (\bibinfo {year} {2018})}\BibitemShut {NoStop}%
\bibitem [{\citenamefont {Marco}\ \emph {et~al.}(2018)\citenamefont {Marco},
  \citenamefont {Valtolina}, \citenamefont {Matsuda}, \citenamefont {Tobias},
  \citenamefont {Covey},\ and\ \citenamefont {Ye}}]{DeMarco2018}%
  \BibitemOpen
  \bibfield  {author} {\bibinfo {author} {\bibfnamefont {L.~D.}\ \bibnamefont
  {Marco}}, \bibinfo {author} {\bibfnamefont {G.}~\bibnamefont {Valtolina}},
  \bibinfo {author} {\bibfnamefont {K.}~\bibnamefont {Matsuda}}, \bibinfo
  {author} {\bibfnamefont {W.~G.}\ \bibnamefont {Tobias}}, \bibinfo {author}
  {\bibfnamefont {J.~P.}\ \bibnamefont {Covey}}, \ and\ \bibinfo {author}
  {\bibfnamefont {J.}~\bibnamefont {Ye}},\ }\href@noop {} {\bibfield  {journal}
  {\bibinfo  {journal} {ArXiv e-prints}\ } (\bibinfo {year} {2018})},\ \Eprint
  {http://arxiv.org/abs/1807.00028} {arXiv:1807.00028 [physics.atom-ph]}
  \BibitemShut {NoStop}%
\bibitem [{\citenamefont {Parker}\ \emph {et~al.}(2015)\citenamefont {Parker},
  \citenamefont {Dietrich}, \citenamefont {Kalita}, \citenamefont {Lemke},
  \citenamefont {Bailey}, \citenamefont {Bishof}, \citenamefont {Greene},
  \citenamefont {Holt}, \citenamefont {Korsch}, \citenamefont {Lu},
  \citenamefont {Mueller}, \citenamefont {O'Connor},\ and\ \citenamefont
  {Singh}}]{Parker2015}%
  \BibitemOpen
  \bibfield  {author} {\bibinfo {author} {\bibfnamefont {R.~H.}\ \bibnamefont
  {Parker}}, \bibinfo {author} {\bibfnamefont {M.~R.}\ \bibnamefont
  {Dietrich}}, \bibinfo {author} {\bibfnamefont {M.~R.}\ \bibnamefont
  {Kalita}}, \bibinfo {author} {\bibfnamefont {N.~D.}\ \bibnamefont {Lemke}},
  \bibinfo {author} {\bibfnamefont {K.~G.}\ \bibnamefont {Bailey}}, \bibinfo
  {author} {\bibfnamefont {M.}~\bibnamefont {Bishof}}, \bibinfo {author}
  {\bibfnamefont {J.~P.}\ \bibnamefont {Greene}}, \bibinfo {author}
  {\bibfnamefont {R.~J.}\ \bibnamefont {Holt}}, \bibinfo {author}
  {\bibfnamefont {W.}~\bibnamefont {Korsch}}, \bibinfo {author} {\bibfnamefont
  {Z.-T.}\ \bibnamefont {Lu}}, \bibinfo {author} {\bibfnamefont
  {P.}~\bibnamefont {Mueller}}, \bibinfo {author} {\bibfnamefont {T.~P.}\
  \bibnamefont {O'Connor}}, \ and\ \bibinfo {author} {\bibfnamefont {J.~T.}\
  \bibnamefont {Singh}},\ }\href {\doibase 10.1103/PhysRevLett.114.233002}
  {\bibfield  {journal} {\bibinfo  {journal} {Phys. Rev. Lett.}\ }\textbf
  {\bibinfo {volume} {114}},\ \bibinfo {pages} {233002} (\bibinfo {year}
  {2015})}\BibitemShut {NoStop}%
\end{thebibliography}%

\clearpage

\end{document}